\documentclass[10pt,letterpaper]{article}
\usepackage{opex3}
\usepackage{epstopdf}
\usepackage{mathrsfs}
\usepackage{amssymb}
\usepackage{cite}
\begin{document}
\title{Statistics and control of waves in disordered media}
\author{Zhou Shi,$^{1,2*}$, Matthieu Davy,$^{3}$ and Azriel Z. Genack $^{1,2}$}
\address{$^1$Department of Physics, Queens College of the City University of New York, \\Flushing, New York, 11367, USA\\
$^2$Graduate Center, City University of New York, New York, NY 10016, USA\\
$^3 $Institut d'Electronique et des T\'{e}l\'{e}communications de Rennes, \\University of Rennes 1,Rennes, 35042, France}
\email{$^*$zhou.shi@qc.cuny.edu} 
\begin{abstract}
Fundamental concepts in the quasi-one-dimensional geometry of disordered wires and random waveguides in which ideas of scaling and the transmission matrix were first introduced are reviewed. We discuss the use of the transmission matrix to describe the scaling, fluctuations, delay time, density of states, and control of waves propagating through and within disordered systems. Microwave measurements, random matrix theory calculations, and computer simulations are employed to study the statistics of transmission and focusing in single samples and the scaling of the probability distribution of transmission and transmittance in random ensembles. Finally, we explore the disposition of the energy density of transmission eigenchannels inside random media.\end{abstract}

\section{Introduction}
The scattering of waves by disorder is more the rule than the exception in condensed materials. Scattering is the source of electrical resistance and makes materials opaque to classical waves. Whereas, all ordered samples of a given material are alike, every disordered sample is disordered in its own way. This can be seen directly in the distinct grainy pattern of intensity of scattered laser light for each sample \cite{1975a}. The speckle pattern within or beyond an illuminated sample arises from the interference of waves following all possible paths to any point. Though the wave field appears to be completely random in disordered samples, it is still possible to manipulate the incident wavefront to control the net transmission and to focus the transmitted wave. This was first achieved by focusing light to a point utilizing feedback from the transmitted wave to manipulate the input wave \cite{2007d}. In principle, a considerable measure of control over the output wave can be exerted by tailoring the incident wave if the field transmission coefficients between all incident and outgoing channels are known \cite{2010c}. These coefficients are the elements of the transmission matrix (TM). In this review, we consider the use of the TM to describe the propagation and control of classical waves in random media. This will be placed in the electronic context in which the TM in the quasi-one-dimensional (quasi-1D) wire geometry was first utilized by Fisher and Lee \cite{1981c} to describe electronic conductance and by Dorokhov to describe the scaling of the eigenvalues of the transmission matrix and the conductance \cite{1982a,1984a}. Because of the strong analogy between electronic and classical wave transport, it is possible to demonstrate fundamental predictions for electron transport in equivalent settings for classical waves in cases in which the predictions have not been demonstrated for electrons. In addition, consideration of the TM for classical waves has stimulated the investigation of universal aspects of transmission eigenvalues such as the dynamics and spatial profiles of the energy density of transmission eigenchannels, which cannot be explored directly in conductance measurements. 

The TM is a powerful tool for studying transport in disordered media in which the random field pattern excited by a monochromatic wave is coherent in time throughout the sample \cite{1991b,2007e,2010a}. For classical waves, thermal fluctuations in position of the sample's constituents in static samples are much smaller than the wavelength so that the wave may be temporally coherent even on a macroscopic scale. In contrast, the electron wave is only phase-coherent in micron-sized conductors at ultralow temperature. Such samples lie between the microscopic atomic scale and macroscopic lengths. The suppression of diffusion \cite{1958b} and the enhancement of fluctuations in conductance \cite{1985b,1985c,1985d} due to interference in samples of intermediate or mesoscopic dimensions have been the focus of studies of electronic transport. 

Anderson \cite{1958b} showed that electrons in an unbounded lattice would be exponentially localized when the range of random values of the electron energy at a site is sufficiently large even when they are not trapped energetically. This leads to the absence of electron diffusion. Unlike the speckle pattern that is washed out by averaging over disorder, Anderson localization is an interference effect that survives averaging over random configurations. The suppression of average transport is a consequence of the constructive interference of partial waves that follow the same path returning to a coherence volume in the sample but in time-reversed order \cite{1969a,1984c}, as illustrated in Fig. 1. The transverse dimensions of the coherence volume is the field correlation length $\lambda/2$. Because the amplitudes and phases associated with such time-reversed partial waves are identical, the amplitude for return for paths following the loop for both senses is twice that for a single partial wave following the loop in either direction. The probability of return of a particle is therefore four times that for a single path and thus twice the probability given by the incoherent sum of probabilities for the two partial waves of each loop. This is the basis of weak localization \cite{1991b} which reduces average transport. When the probability of return to a coherence volume reaches unity, the wave becomes localized. The effect of weak localization is observed directly as coherent backscattering in the domain of the reflected wavevector \cite{1985g,1985h,1986f}. Reflection is enhanced by a factor of two over the background level in the retroreflection direction. The angular distribution of backscattered light is the Fourier transform of the intensity distribution of the point spread function for light incident on the incident surface. The enhancement falls in an angular range of $1/k\ell$, where $k$ is the wavenumber and $\ell$ the transport mean free path. Measurements of coherent backscattering therefore allow for the determination of the mean free path.
\begin{figure}[htc]
\centering
\includegraphics[width=2in]{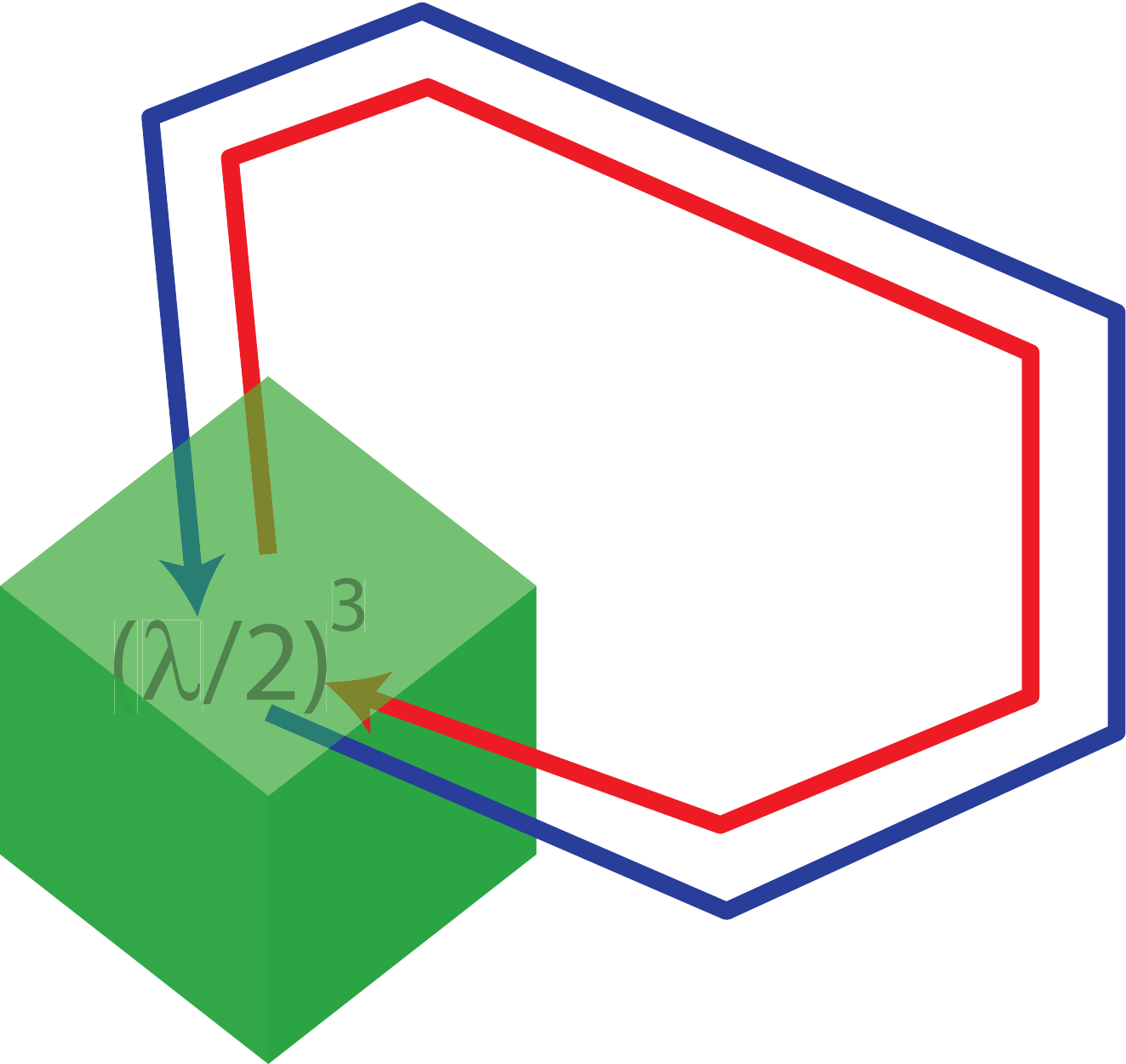}
\caption{Illustration of a wave returning to a coherent volume of $(\lambda/2)^3$ in three dimensions and its time-reversed partner.} \label{Fig1}
\end{figure}

Ioffe and Regel \cite{1960b} suggested that, an electron cannot be considered to be freely propagating if the scattering mean free path were only a fraction of the wavelength. This yields the localization criterion in three dimensions, $k\ell<1$. This condition is also given by a calculation of the return probability based on diffusion theory in three dimensions. Subsequently, Thouless \cite{1974c,1977a} argued that localization in any dimension should depend only upon the dimensionless ratio of the average width and spacing of energy levels, which has come to be known as the Thouless number, $\delta = \delta E/\Delta E$. Electronic levels correspond to quasi-normal modes or resonances for classical waves in open systems. The width of levels, which in units of angular frequency is equal to the leakage rate of energy from the sample, is closely linked to the sensitivity of level energies to changes at the boundary since both are proportional to the ratio of the strengths of the mode at the boundary relative to the interior of the sample. When the wave is localized within the interior of a sample, the amplitude of the wave is exponentially small at the boundary and the mode is weakly coupled to the surrounding medium. Its lifetime is then long and its linewidth correspondingly narrow, so that, $\delta E<\Delta E$. On the other hand, when the wave is diffusive, modes extend throughout the medium; the wave then couples readily to its surroundings and the width of a typical level overlaps several modes, $\delta E>\Delta E$. Thus, the localization threshold occurs at $\delta=1$.

Thouless \cite{1974c,1977a} showed that for diffusive samples, $\delta$ is equal to the conductance in units of the quantum of conductance $\delta={\textsl g}\equiv G/(e^2/h)$. Since the level width falls inversely with the square of sample length for diffusive samples while the spacing between levels falls only inversely with the sample length in the quasi-1D wire geometry of constant cross section, $\delta$ falls inversely with sample length. Thus there will always be a crossover from diffusive to localized wave transport as the length of a wire increases. The crossover will occur at a sample length $L = \xi = N\ell$, at which $\delta={\textsl g}=1$ \cite{1977a}. Here, $\xi$ is the localization length of the sample and $N$ is the number of channels. The channels can be the propagating transverse modes in the ideal leads attached to a conductor at a given voltage. The total number of channels is approximately $2\pi A/\lambda^2$ including two electronic spin states, where $A$ is the constant cross sectional area of the sample and $\lambda$ is the electron wavelength. Abrahams {\it et al.} \cite{1979a} argued that the variation of average {\textsl g} with the dimensions of the sample should depend only upon the value of {\textsl g} and the dimensionality. A localization transition is predicted to exist only in 3D in which {\textsl g} falls essentially exponentially with increasing sample size for ${\textsl g}<1$ and increases for ${\textsl g}>1$ \cite{1979a}. 

Since the conductance exhibits large fluctuations for localized waves, the mean value of the conductance is insufficient to characterize the nature of electronic transport and the full distribution of the conductance needs to be given. Anderson {\it et al.} \cite{1980a} hypothesized that the logarithm of the conductance of localized waves in one-dimensional (1D) samples follows a Gaussian distribution with a variance approaching the average of the logarithm of the conductance, $\langle \ln {\textsl g}\rangle=-L/\xi$. The scaling of the distribution of conductance for localized waves in 1D would then be completely specified by the single parameter $L/\xi$. 

Electronic conductance in disordered 1D conductors and transmission of classical waves in mesoscopic media are seen to have equivalent descriptions in the 1D Landauer relation, ${\textsl g}=T$, where $T$ is the transmission \cite{1970a,1981b}. The Landauer relation was extended to multichannel mesoscopic systems via the TM \cite{1981c}. The elements $t_{ba}$ of the transmission matrix $t$ are the field transmission coefficients between incoming channels {it a} and output channels {\it b}. The dimensionless conductance in quasi-1D samples is equal to the sum of $N^2$ pairs of intensity transmission coefficient, which is the transmittance, $T$, $g=\sum_{a,b}^N |t_{ba}|^2\equiv T $. The dimensionless conductance or transmittance can also be expressed as the sum of the transmission eigenvalues, $g=\sum_{n}^N \tau_n$. Here, the $\tau_n$ are the eigenvalues of the matrix product $tt^\dagger$. The transmission eigenvalues can also be obtained via the singular value decomposition of the TM, $t=U\Lambda V^\dagger$, where $U$ and $V$ are unitary matrices with columns which are the singular vectors yielding the output and input of the transmission eigenchannels and $\Lambda$ is the diagonal matrix with elements $\sqrt{\tau_n}$.

Dorokhov \cite{1982a,1984a} considered the scaling of {\textsl g} in samples made up of $\it N$ parallel 1D disordered chains with transverse coupling to create a quasi-1D sample. He found that, instead of a single localization length $\xi$, there exists $\it N$ auxiliary localization lengths, $\xi_n$, which describe the scaling of the averages of each of the transmission eigenvalues vs. length via the relation, 
\begin{equation}
\tau_n=1/\cosh^2(L/\xi_n). 
\end{equation}
The inverse auxiliary localization lengths are   are given by $1/\xi_{n}=(2n-1)/2\xi$ for integers $n<N/2$ so that these are equally spaced with spacing $1/\xi_{n+1}-1/\xi_n=1/\xi$ \cite{1982a,1984a,1990g,1995f, 1997c,2012f,2014a}. 

An intuitive Coulomb gas model \cite{1990g,1991c} was developed to describe the statistics of transmission eigenvalues. This is based on the model originally introduced by Dyson \cite{1962a} to visualize the logarithmic repulsion between eigenvalue of a large random Hamiltonian matrix \cite{1958a}. In the Coulomb gas model, the $\tau_n$ are associated with the positions of parallel lines of charges at $x_n=L/\xi_n=(2n-1)L/2\xi)$ for $n<N/2$ and image charges of the same sign at $-x_n$ \cite{1991c}. The line charges in the model are embedded in a compensating continuous charge background of opposite sign. The background charge tends to screen the line charges from the interaction with other line charges. For diffusive waves, the spacing between neighboring charges is small compared to the screening length. As a result, there exists a logarithmic repulsion between the transmission eigenvalues due to the Coulomb interaction. The roughly uniform distribution of $x_n$ imposed by the mutual repulsion of charges leads to the bimodal distribution of transmission eigenvalues $\tau$ with peaks in the probability distribution for $\tau$ near unity and at exponentially small values of $\tau$, $\rho(\tau) = \frac{{\textsl g}}{2\tau\sqrt{(1-\tau)}}$ \cite{1984a,1988c, 1990f,1994h,1997c}. The repulsion between charges for diffusive waves leads to universal conductance fluctuations. In quasi-1D disordered systems with time-reversal symmetry, ${\rm var}(T)=2/15$ \cite{1997c}. 

Universal conductance fluctuations were first measured in mesoscopic conductors at low temperatures \cite{1985b,1986e}. Fluctuations of conductance of a gold wire at 10 mK as a function of magnetic field applied perpendicular to the wire are shown in Fig. 2. Dorokhov \cite{1984a} showed that transport through disordered conductors is dominated by {\textsl g} eigenvalues with $\xi_n>L$ and so transmission through these channels is of order unity. The number of such open eigenchannels with $\tau_n>1/e$, which is essentially the conductance for diffusive quasi-1D samples, is inversely proportional to the sample length in accord with Ohm's law. For localized waves, however, the spacing between charges in the Coulomb gas model is greater than the screening length so that the Coulomb interaction between charges is diminished. The smallest of the $x_n$, $x_1$, is larger than unity for localized waves so the transmission eigenvalues fall exponentially with $n$, $\langle \tau_{n+1}\rangle/\langle \tau_n\rangle = \exp(-2L/\xi)$. As a result, the conductance is dominated by the highest transmission eigenchannel $\tau_1$ \cite{1984a,1990g,1991c,1995f,2012f,2014a}. 
\begin{figure}[htc]
\centering
\includegraphics[width=3in]{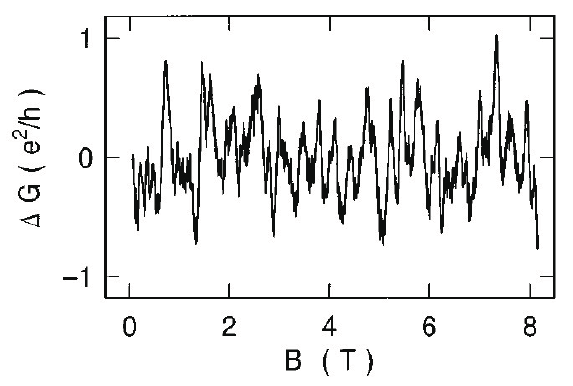}
\caption{Fluctuations of conductance of a gold wire at 10 mK as a function of magnetic field applied perpendicular to the wire. The wire is of length 310 nm and width 25 nm. (From Ref. \cite{1986e})} \label{Fig1}
\end{figure}

In contrast to electronic systems, for which only the conductance can be measured, it is in principle possible to measure the elements of the TM for classical waves. Thus the transmission coefficients for the field, $t_{ba}$, intensity $T_{ba}=|t_{ba}|^2$, total transmission $T_a=\sum_{b}^{N} T_{ba}$ and transmittance $T$ can be measured. Spectra of these variables for a single sample supporting localized and diffusive waves in different frequency ranges in which {\textsl g} =6.9 and {\textsl g} = 0.37 are shown in Fig. 3. The statistics of random ensembles of statistically equivalent samples can be obtained in microwave measurements for a succession of sample realizations produced by briefly rotating a copper tube containing randomly positioned dielectric spheres. Ensembles of random system can be studied optically by illuminating different regions of a random slab. The statistics of internal motion of colloidal samples can also be explored in measurements of the temporal correlation of transmitted light \cite{1987b,1988e,1998a}. 
\begin{figure}[htc]
\centering
\includegraphics[width=4in]{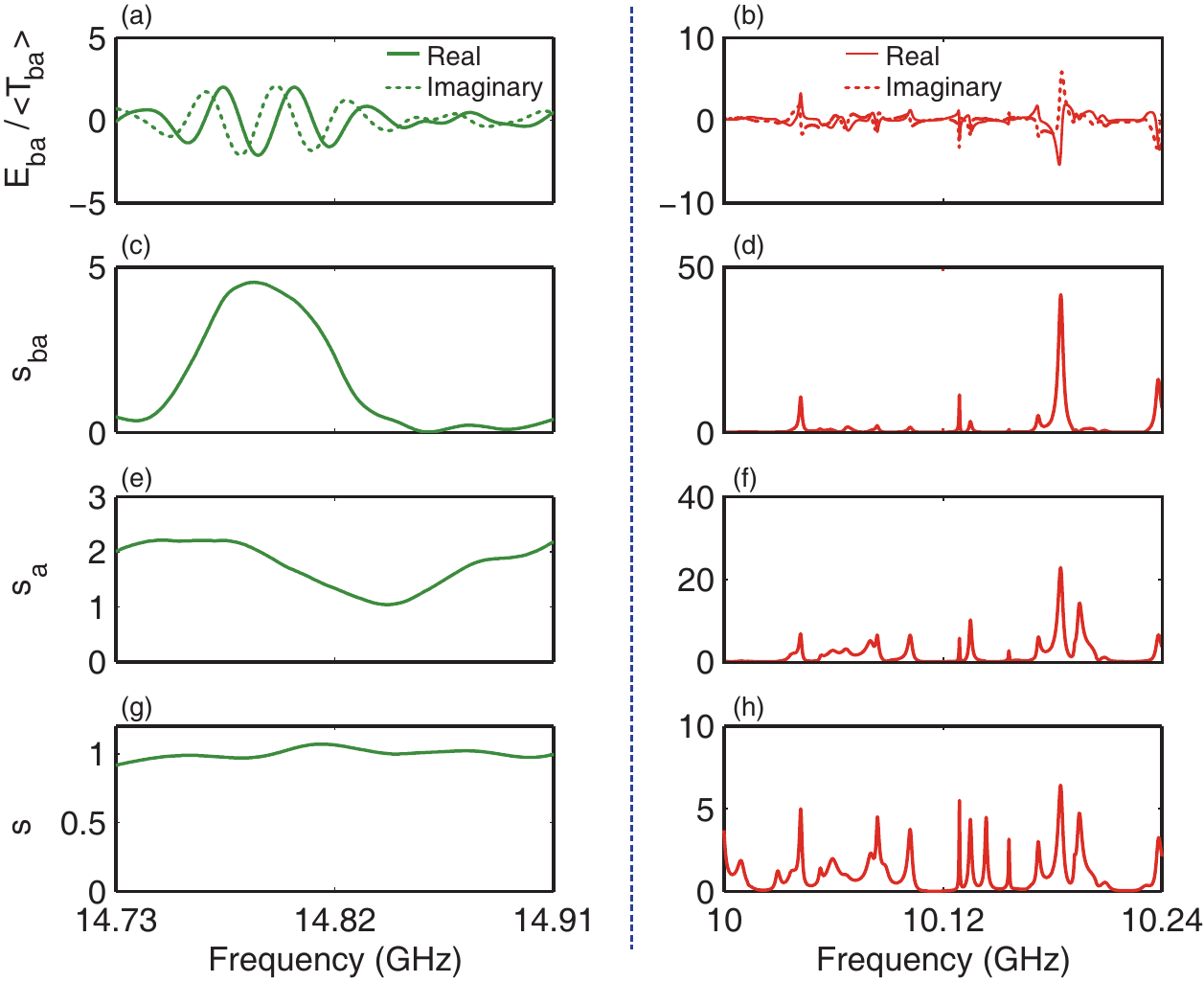}
\caption{Spectra of the transmission coefficients normalized to ensemble average values. In- and out-of-phase field transmission coefficients, field $E_{ba}/\sqrt{\langle T_{ba}/\rangle}$, transmission coefficients of intensity $s_{ba}=T_{ba}/\langle T_{ba}\rangle$, total transmission $s_a=T_a/\langle T_a\rangle$ and transmittance $s=T/\langle T\rangle$ for microwave radiation propagating through a random waveguides for a diffusive sample (left column) and a localized sample (right column). For diffusive waves, $\delta>1$, many modes contribute to transmission at all frequencies and for all source and detector positions. In contrast, for localized waves $\delta<1$, distinct peaks appear when the incident radiation is on resonance with a quasi-normal mode. The resonance condition holds for all source and detector positions and therefore sharp peaks remain even when transmission is integrated over space.} \label{Fig1}
\end{figure}

The first question that was addressed in bounded samples was the scaling of conductance. This is practically difficult to realize in electronic measurements because of the limited scale of mesoscopic samples and the difficulty of producing statistically equivalent samples. The scaling of optical transmission in the slab geometry can be measured in the slab geometry by forming a wedge of random material and focusing a laser beam on the wedge as it is translated perpendicular to its vertex \cite{1987a, 1997d}. The transmission of light through a slab of rutile titania particles in a polystyrene matrix is seen in Fig. 4. The lower envelopes of the transmission curves in Figs. 4a,b fall inversely with sample thickness $L$ up to a length equal to the absorption length $L_a$ indicating that the wave is diffusive. The absorption length is given by $L_a=\sqrt{D\tau_a}$, where the diffusion coefficient is $D=v_e\ell/3$, $v_e$ is the energy velocity in the medium and $\tau_a$ is the absorption time. Spikes in transmission occur at spots in the sample at which a pit has been gouged out of the otherwise smooth surface in the polishing process. Spectra of transmitted intensity in the far field at different sample thicknesses are shown in Fig. 4(c). The correlation frequencies of fluctuations in transmission at different sample thicknesses give the average linewidth of electromagnetic modes of the medium \cite{1990a}. In this diffusive sample, the correlation frequency scales as $L^{-2}$ for $L<L_a$ \cite{1990a}. The field correlation function and the temporal profile of transmission form a Fourier transform pair and give the diffusion coefficient and absorption length which is consistent with measurement of the scaling of static transmission \cite{1990a}. 
\begin{figure}[htc]
\centering
\includegraphics[width=4.5in]{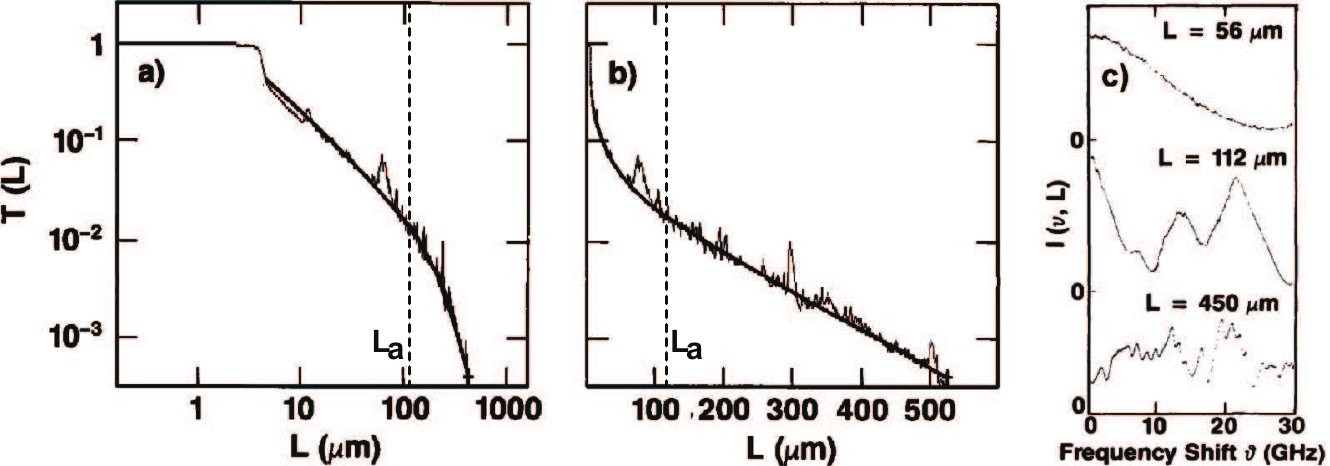}
\caption{Scaling of optical transmission through a wedged sample consisting of TiO$_2$ embedded in polystyrene. (a,b) Log-log and semi-log plots of the scaling of the transmission coefficient, $T(L)$. The inverse variation of $T(L)$ with $L$ is seen in the log-log plot and the exponentially falloff of transmission beyond the absorption length $L_a$ is seen in the semi-log plot. The absorption length $L_a=112 \pm 5\mu m$ of the sample is indicated by the dashed line in both figures. The solid line is a fit of the expression $T(L)=5\alpha D/v\sinh(\alpha L)$ to the data. Here $D$ is the diffusion coefficient, $v$ is the speed of light in the medium and $\alpha$ is the inverse of $L_a$. (c) Spectra of transmission for various sample lengths. The measured transmission in these experiments never reaches zero because the polarized intensity only vanishes at point singularities while measurements are made with a finite aperture placed before the photomultiplier tube. (From Ref. \cite{1987a})} \label{Fig2}
\end{figure}

Establishing that a wave is localized through scaling is a challenge in the presence of absorption of classical waves and of dephasing of the electronic wave function \cite{1984b,1991e,1993a,1997d,1999a,2000a}. But the magnitude of fluctuations of intensity and total transmission normalized by their ensemble average values give a straightforward measure of the degree of localization which is robust against absorption \cite{2000a}. The magnitudes of relative fluctuations in electronic conductance \cite{1985b,1985c,1985d}, classical intensity and total transmission and the degrees of spatial, spectral and temporal correlation of flux increase through the localization transition \cite{1986b,1987d,1988a,1988b,1988d,1990a,1990c,1990d,1998a,2002a,1989a,1994b,1995b,1995a,1996c,1997b,1999b}. 
Large fluctuations in transmitted intensity were observed in the critical regime of the Anderson localization transition in acoustic measurements in a slab of brazed aluminum beads \cite{2008a}. These results are consistent with the agreement found between the non-exponential decay of a transmitted pulse measured in this sample and self-consistent localization theory \cite{2006b}. Microwave measurements in quasi-1D samples of the time-dependence of transmission in more deeply localized samples \cite{2009c} show dramatically reduced decay rates for transmission compared to both diffusion theory and the self-consistent localization theory. Since the slow-down of decay rates is associated with long-lived modes \cite{2000c,2003a,2011j}, this indicates that the self-consistent theory is valid near the localization transition but not for deeply localized waves \cite{2006b,2010e,2010f,2014g}. The approach to localization can also be seen in the saturation of the spread of a wave with time delay on the output surface of the sample, which is independent of absorption. This was observed in acoustic measurements \cite{2008a} and in pulsed laser experiments in random slabs of rutile titania particles \cite{2013m}. The local strength of scattering in the optical experiments was determined from measurements of the width of the coherent backscattering cone \cite{2006c,2013m}. Values of the product of the wave vector and the transport mean free path, $k\ell$ were larger than the value predicted by the Ioffe-Regel criterion \cite{1960b} for localization in three dimensions of $k\ell<1$. The lowest value of $k\ell$ for these samples was 2.5. The Ioffe-Regel criterion is consistent with the Thouless criterion for localization \cite{1977a,1990h}. Localization in samples with $k\ell>1$ might occur in random samples with residual order \cite{1987e} or in samples with resonances in individual scattering elements \cite{1989f,1991h,1996d,2001k}. In such samples, the density of states (DOS) may fall below the uniform medium value $8 \pi \nu^2/v_p^3$, where $\nu$ is the frequency, $v_p$ is the phase velocity. Localization might then occur from values of $k\ell>1$.

Additional evidence that these samples are in the critical regime is seen in the continuous slowing decay of pulsed transmission in these samples. This shows that the exponential decay of transmission with time delay expected for diffusive samples breaks down at long times. Departures from an exponential decay of transmission for diffusive waves with values of ${\textsl g}\sim2.5$ have been observed in microwave transmission \cite{2003a}. This indicates that the distribution of mode lifetimes increases even in diffusive samples as the crossover to localization is approached \cite{2000c,2003a}. An important question is whether the time dependence of the transverse spatial profile represent the reduced spatial extent of some pre-localized modes as the localization transition is approached or localization \cite{1995g,2000c,2002e}. Transverse optical localization \cite{1989e} has been observed in samples that are uniform in the longitudinal direction but disordered in two \cite{2007a} or one \cite{2008e} transverse dimension. Whether it is possible to localize electromagnetic radiation in three dimensions in samples without residual order remains an open question \cite{1987e,2014j}.

The TM has been recently measured for light \cite{2010c,2012c,2012g,2013g}, microwave radiation \cite{2012f} and acoustic waves \cite{2009b,2014d} in static samples. The TM gives a full description of the degree to which the incident wavefront can be manipulated to control the transmitted wave \cite{2008b,2010c}. This may be exploited to focus radiation and enhance or suppress transmission for applications in imaging and telecommunications \cite{2010c,2008g,2010b,2010g,2010i,2011g,2011h,2012a,2012b,2012c,2012d,2012f,2012g,2013g,2012i,2012j,2013e,2013f,2013h,2013i,2014b,2014c,2014h,2014k,2014l}. The ability to control propagation of classical waves in a random medium was first demonstrated in acoustics by focusing a pulse within and through a scattering medium \cite{1992c}. This is achieved by time-reversing the signal recorded by an array of transducers. The measured time-varying signal is played back in time and a pulse emerges at the location of the source. For monochromatic radiation, it is possible to focus the transmitted wave by shaping the incident waveform. Vellekoop and Mosk \cite{2007d} demonstrated optical focusing through an opaque sample by adjusting the incident wavefront reflected from a spatial light modulator using a genetic algorithm with feedback from the intensity at the focus. When the TM of the scattering sample is determined, it is possible to focus at any desired channel on the output surface. This can be achieved by producing an incident waveform $t^*_{\beta a}$, which is the phase conjugate of the field one would obtain on the incident plane when a source is placed at a point $\beta$ on the output plane. This would bring the fields from different incident channels $a$ arriving at $\beta$ in phase so that the intensity at the point is greatly increased by constructive interference. Vellekoop and Mosk \cite{2008b} found that the background intensity in the transmitted wavefront for optimally focused radiation is also enhanced relative to the average transmitted intensity. In addition, the total transmission can be significantly enhanced to give nearly complete transmission when the incident wavefront corresponds to the highest transmission eigenchannel. 

The TM gives the fullest account of transmission and conductance and enables the control of the transmitted flux. However, the TM cannot provide the intensity profiles of the transmission eigenchannels inside the sample. Recent simulations by Choi {\it et al.} \cite{2011e} explored the spatial profile of the intensity of eigenchannels inside a single random sample. They have observed that the energy stored within high transmission eigenchannels is enhanced. This was confirmed by G\'{e}rardin {\it et. al.} \cite{2014d} in measurements of flexural waves propagating in a disordered elastic waveguide. Using a laser source and a heterodyne interferometer, they have measured the full scattering matrix of the medium. After a proper normalization of the measured scattering matrix, they found that the transmission eigenvalues follow a bimodal distribution in agreement with theoretical prediction. They have also demonstrated that the intensity associated with high transmitting eigenchannels increases within the medium with a peak located near the middle of the sample. This was presented in Fig. 5. We have shown recently in microwave measurements that the integral of the intensity inside the sample for a specific eigenchannel yields the dwell time of the eigenchannel \cite{2015a}. The dwell time of the each of the eigenchannel is also equal to its contribution to the DOS of the sample so that the sum of the dwell times for $N$ eigenchannels gives the DOS of the disordered material. 
\begin{figure}[htc]
\centering
\includegraphics[width=4.5in]{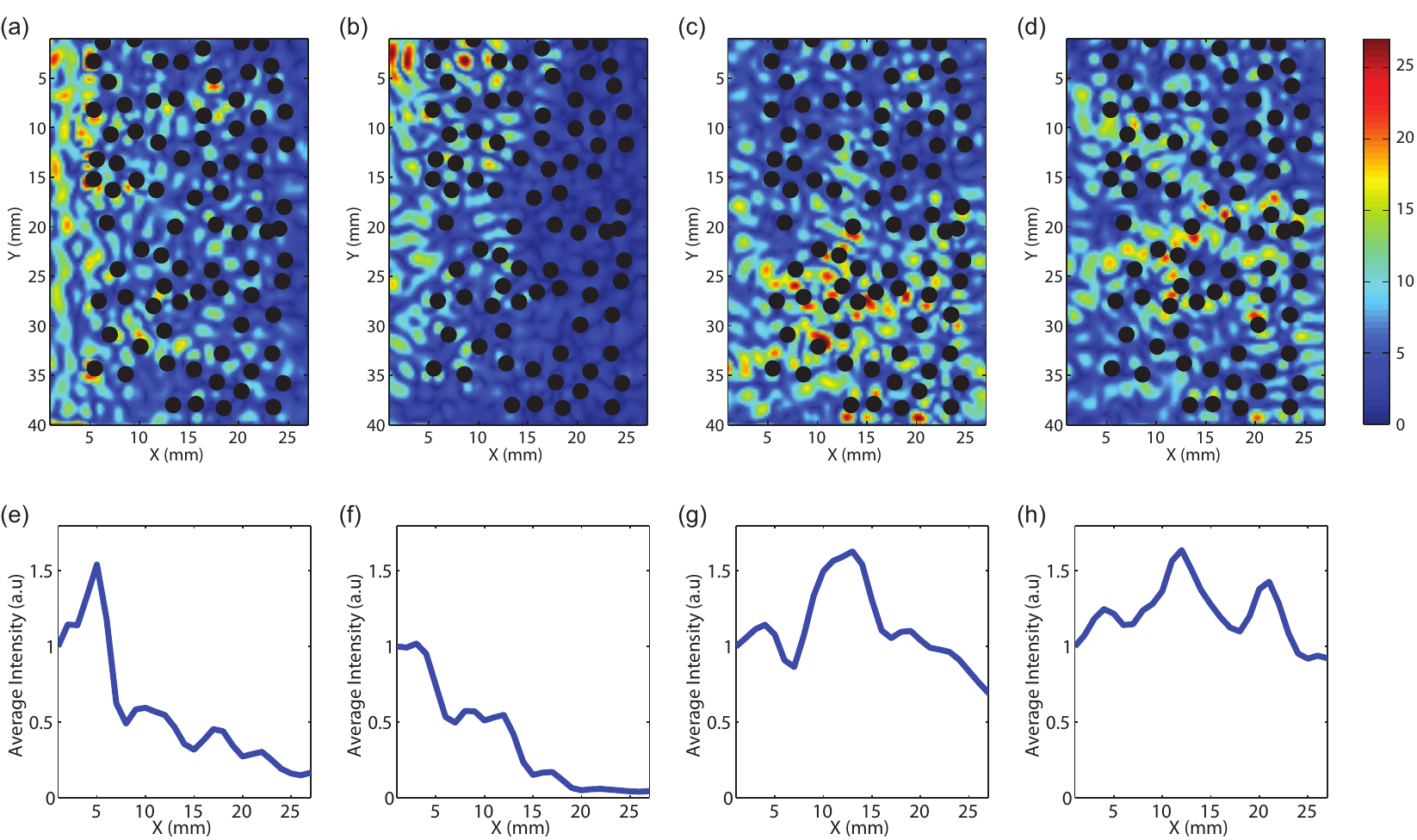}
\caption{Absolute value of the wave field in the scattering medium at f = 0.36 MHz associated with (a), an incident plane wave, (b), a closed eigenchannel, (c), an open eigenchannel deduced from the measured \textbf{S} matrix, and (d), an open eigenchannel deduced from the normalized matrix $\hat{\textbf{S}}$. The corresponding intensities averaged over the wave guide section (y axis) are shown versus depth x in lower panels (e)–(h). They are all normalized by the intensity at the plane of sources (x = 0). (From Ref. \cite{2014d})} \label{Fig3}
\end{figure}

The average intensity falls linearly inside the medium in diffusive samples, as required by the Fick's law. For a localized sample, wave interference cannot be neglected in calculating the average profile \cite{2008c,2008d}. The intensity profile for localized waves excited by radiation incident from one side of the sample is found to fall more rapidly toward the center of the open medium \cite{2010e,2010f,2014g}. This profile can be found by solving a generalized diffusion equation with a position dependent diffusion coefficient, which reflects the increasing renormalization of transport with depth into the sample due to wave interference \cite{2010e}. Recent simulations of intensity profiles inside the material for a random ensemble have suggested a universal expression for the profile of the energy density of the transmission eigenchannels \cite{2015b}. These profiles may be expressed in terms of the auxiliary localization lengths of the corresponding transmission eigenchannels. The structure of these profiles is given by solutions of the generalized diffusion equation \cite{2015b}. 

In the sections of this review that follow, we will focus on recent microwave measurements and computer simulations of spectra of the TM in quasi-1D random waveguides in the crossover from diffusive to localized waves. In Section II, we discuss the statistics of transmission in single sample realizations and the statistics of the transmittance in ensembles of random samples. The statistics in singe samples with given transmittance $T$ are found to depend upon only a single parameter, which is the eigenchannel participation number, $M\equiv (\sum_{n}^{N} \tau_n)^2/\sum_{n=1}^{N} \tau_n^2$ \cite{2013a}. The probability distribution of $T$ is observed to change from a one-sided log-normal distribution to a log-normal distribution as $M\rightarrow1$ as the wave becomes deeply localized. In addition, the statistics of transmission are found to approach predictions for 1D. In particular, the single parameter scaling (SPS) hypothesis predicted for 1D random samples is found experimentally to hold for deeply localized waves in a quasi-1D geometry. Focusing of both static and pulsed radiation through and within opaque samples is discussed in Section III. Expressions for the spatial profile of the focused beam and the contrast between the focused and background intensities are demonstrated. In Section IV, we show that a complementary set of parameters to the transmission eigenvalues provides the dwell time and the contribution of each eigenchannel to the DOS. These parameters correspond to the integral of the energy density profile inside the sample for the corresponding transmission eigenchannel. A summary of recent results and prospects for applications are given in in section V.

\section{Statistics of transmission in single samples and ensembles of random samples}
The statistics of transmission in a single disordered sample are of particular interest since applications are typically in a specific sample. Such statistics are an essential part of the statistics of transmission over a collection of random samples. The transmitted field with a given polarization for an incident channel $a$ in a single random speckle pattern is a complex Gaussian random variable. Since the transmission is the square of the field on the output surface, $T_{ba}=|t_{ba}|^2$ the corresponding probability distribution of intensity relative to the average in the speckle pattern, $T_{ba}/(\sum_{b}^{N} T_{ba}/N)=NT_{ba}/T_a$, is a negative exponential function, $P(NT_{ba}/T_a)=\exp(-NT_{ba}/T_a)$ \cite{1985a,2010d}. Thus, the statistics of the transmitted flux relative to its average in a single TM are fully specified by the statistics of total transmission $T_a$ normalized by its mean within the TM, $T_a/\langle T_a\rangle_a=NT_a/T$. Here, $\langle \dots \rangle_a$ indicates the average over all incident channels $a$ in a single sample. 

\begin{figure}[htc]
\centering
\includegraphics[width=3in]{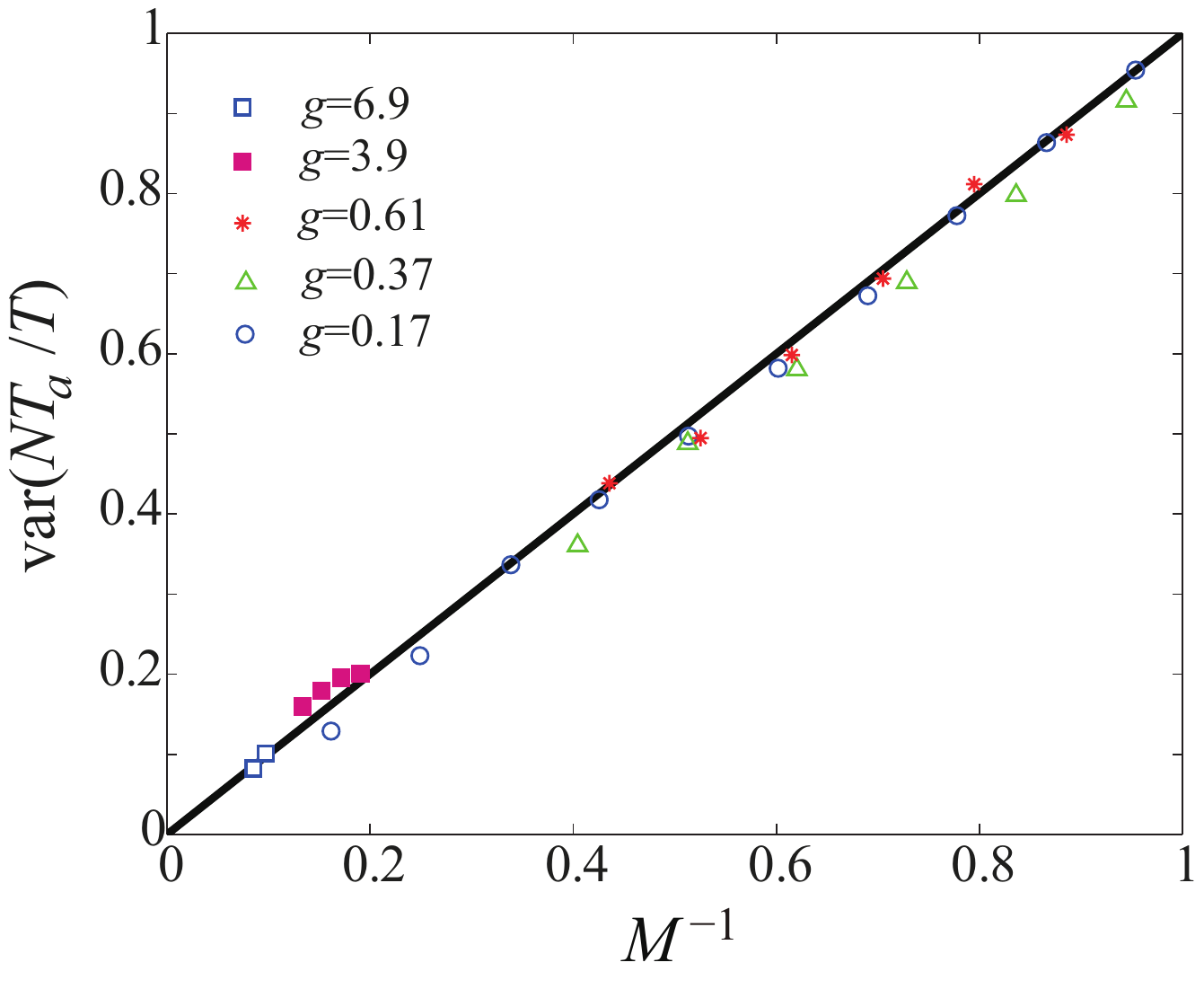}
\caption{The variance of the normalized total transmission vs. the inverse of the eigenchannel participation number $M$ is obtained from measurements by grouping single samples with the same value of $M^{-1}$ from ensembles with a wide range of values of {\textsl g}. The solid line is ${\rm var}(NT_a/T)=M^{-1}$. (From Ref. \cite{2013a})} \label{Fig3}
\end{figure}
Using the singular value decomposition of the TM, the variance of the total transmission for an incident channel {\it a} relative to the average over all possible incident channels can be expressed as \cite{2013a}, 
\begin{equation}
{\rm var}(NT_a/T)=\frac{\sum_{n}^N \tau_n^2}{(\sum_{n}^N \tau_n)^2}=1/M.
\end{equation}
For diffusive samples without absorption, the bimodal distribution of the eigenvalues gives $\langle M\rangle =3{\textsl g}/2$. The value of {\textsl g} for diffusive waves is close to the number of open eigenchannels with transmission eigenvalues greater than $1/e$ \cite{1984a,1986a}. For deeply localized waves, the number of open eigenchannels falls to zero, but $M$ approaches unity since the transmittance is dominated by a single transmission eigenvalue.

These calculations apply to a single large TM and to subsets of samples with the same values of $M^{-1}$. We consider measurements of samples with small value of $N$ and group together subsets of samples with similar values of $M^{-1}$. Figure 6 demonstrates that ${\rm var}(NT_a/T)$ is close to $M^{-1}$ in subsets of samples with given $M^{-1}$. 

The measurements of microwave transmission described in this review were carried out in collections of samples consisting of randomly positioned alumina spheres with refractive index 3.14 within a copper tube. The dielectric spheres were embedded in Styrofoam shells to produce an alumina volume fraction of 0.068. Microwave radiation was launched and detected by wire antennas connected to a vector network analyzer. The orientation of the wire antennas determines the polarization of the field that is detected. The source and detector antennas are mounted on 2D translation stages and moved to positions on a square grid over the open ends of the waveguide at which spectra of the in- and out-of-phase components of the transmitted electric field $t_{ba}$ are collected. Measurements are made over two frequency ranges of 10-10.24 GHz and 14.7-14.94 GHz. The lower frequency range is just above the first Mie resonance of the individual alumina spheres and so the scattering is strong and the wave is localized in a relatively short length of the random waveguide. The wave is diffusive in the higher frequency range. The number of allowed propagating waveguide modes increases with frequency; they are $\sim$30 and $\sim$66 in the centers of the two frequency ranges. Measurements of spectra of the TM in the two frequency ranges are made for sample lengths between 23 and 102 cm so that ensembles over a wide range of {\textsl g} are studied. The impact of absorption on the statistics of transmission is largely removed by Fourier transforming the field spectrum into the time domain and multiplying the time signal by $\exp(t/2\tau_a)$, where $t$ is the time delay and $\tau_a$ is the absorption time \cite{1993a,2000a}.

\begin{figure}[htc]
\centering
\includegraphics[width=4in]{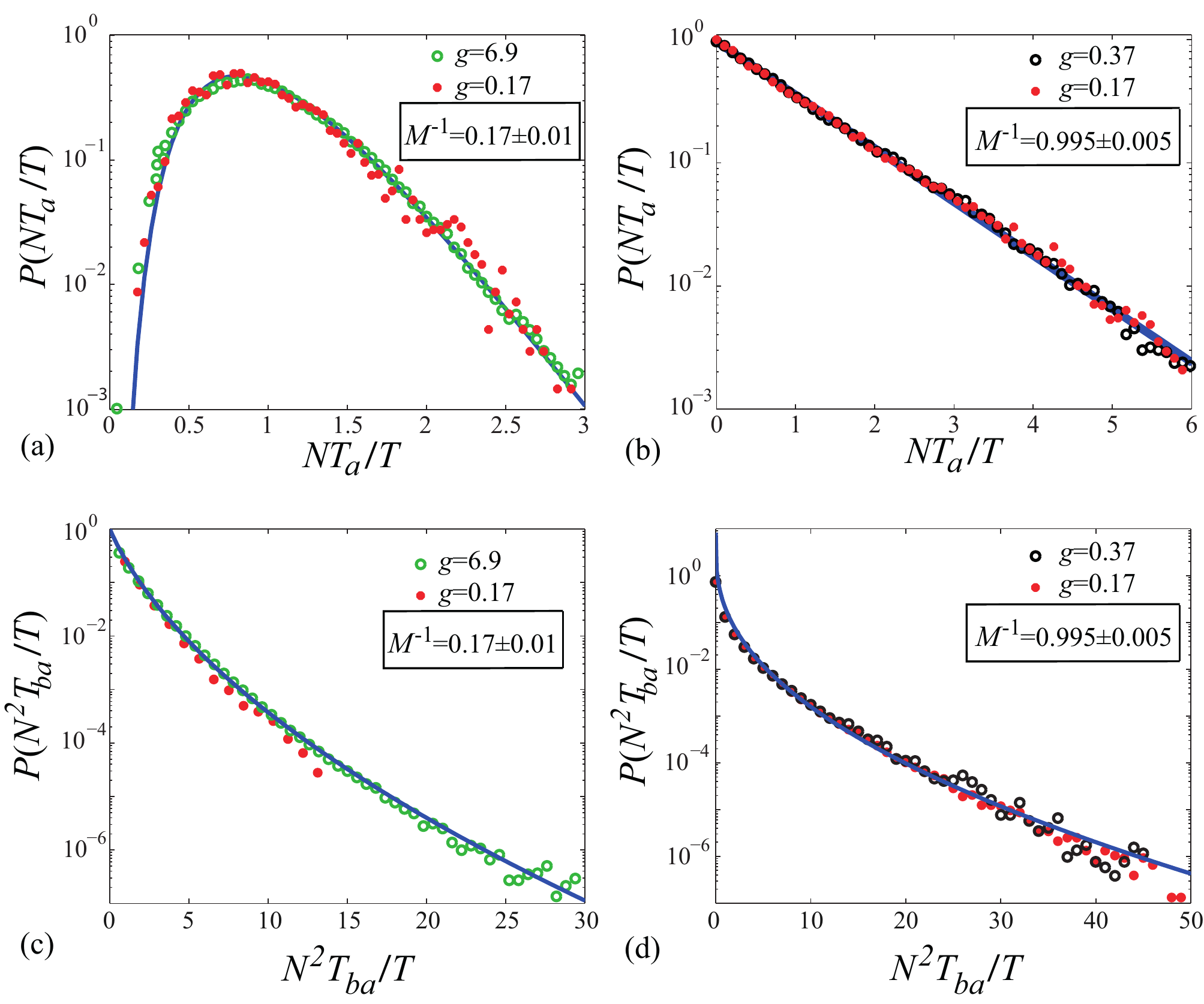}
\caption{Probability distributions of normalized total transmission and intensity in sub-ensembles of random samples with similar values of $M^{-1}$. (a) The probability distributions of normalized total transmission $P(NT_a/T)$ for samples with $M^{-1}=0.17\pm 0.01$ from two random ensembles with {\textsl g} = 3.9 (Green open circles) and {\textsl g} = 0.17 (red filled circles) are seen to overlap, confirming that $P(NT_a/T)$ depends only on the value of $M$. The solid curve is a calculation of $P(T_a/\langle T_a\rangle)$ with ${\rm var}(T_a/\langle T_a\rangle)$ replaced with $M^{-1}$ in the expression from Ref. \cite{1995b,1995a}. (b) $P(NT_a/T)$ for $M^{-1}$ in the range of for two ensembles with ${\textsl g}=0.37$ and 0.17. The solid line is an exponential distribution, $\exp(-NT_a/T)$. (c,d) The normalized intensity distributions, $P(N^2T_{ba}/T)$ for the same as in (a,b). The solid lines are the calculations of the distribution based upon $P(NT_a/T)$ in (a,b) and the exponential falloff of the distribution of the normalized intensity $P(NT_{ba}/T_a)=\exp(-NT_{ba}/T_a)$. (From Ref. \cite{2013a})} \label{Fig4}
\end{figure}
Measured distributions of relative total transmission and intensity for subsets of samples with $M^{-1}$ in the ranges $0.17\pm0.01$ and $0.995\pm0.005$ are shown in Fig. 7. The distributions in samples with the same values of $M^{-1}$ but drawn from ensembles of different {\textsl g} are seen to coincide. This shows that the distributions $P(NT_a/T)$ depend only upon the value of $M$. $P(NT_a/T)$ are also presented in Fig. 7(b) for $M^{-1} $near unity in measurements for two ensembles of localized samples with ${\textsl g}=0.37$ and 0.17. When $\tau_1 \gg \tau_2$, transmission is dominated by the first transmission eigenchannel so that, $t_{ba}=\sum_n u_{bn}\sqrt{\tau_n}v^\dagger_{na}\sim u_{b1}\sqrt{\tau_1}v^\dagger_{1a}$ and $T_{ba}=|u_{b1}|^2\tau_1|v^\dagger_{1a}|^2$. This gives $NT_a/T=|v_{1a}|^2$. The probability distribution $P(NT_a/T)$ is therefore expected to be a negative exponential, $P(NT_a/T)=\exp(-NT_a/T)$. Measurements are in excellent agreement with this conjecture, as shown in Fig. 7(b).

The eigenchannel participation number $M$ can be linked directly to the degree of intensity correlation within a single sample. In the limit $N\gg 1$, infinite-range correlation due to fluctuations in $T$ vanishes if computed for the TM of a single sample. The cumulant correlation function of transmitted intensity relative to its average, $T/N^2$, is then $C_{ba,b^\prime a^\prime}^M=\langle [T_{ba}T_{b^\prime a^\prime}-(T/N^2)^2]/(T/N^2)^2\rangle_M$. The values of residual short-range and long-range correlation determined from the variances of relative intensity and total transmission are unity and $M^{-1}$, respectively, giving 
\begin{equation}
C_{ba,b^\prime a^\prime}=\delta_{a a^\prime}\delta_{b b^\prime}+M^{-1}(\delta_{a a^\prime}+\delta_{b b^\prime}).
\end{equation} 
This leads to ${\rm var}(N^2T_{ba}/T)=1+2M^{-1}$. This is confirmed in measurements by the correspondence of the measured variances of relative intensity for the two values of $M^{-1}$ of 0.17 and 0.995 of 1.38 and 3.04 with the calculated values of 1.34 and 3.00.

Since the statistics of transmission in a single sample at a given frequency depend only on $M$, they do not reflect the nature of transport within a disordered material. This would require knowledge of the distribution of $M$. Previous studies have shown that the crossover to Anderson localization can be tracked in terms of the fluctuation of the intensity and total transmission over collections of samples in an ensemble \cite{1994b,1995a,1995b,1997b,1999b,2000a}. In particular, for diffusive waves in the quasi-1D geometry, the variances of $T_{ba}$ and $T_a$ normalized by their ensemble averages are linked to the dimensionless conductance {\textsl g} via the relation, ${\rm var}(s_{ba}=T_{ba}/\langle T_{ba}\rangle)=1+4/3{\textsl g}$ and ${\rm var}(s_a=T_a/\langle T_a\rangle)=2/3{\textsl g}$, respectively. ${\rm var}(s_a)$ is also equal to the degree of long range intensity correlation, $\kappa\equiv \langle \delta s_{ba}\delta s_{b^\prime a^\prime}\rangle$,where $\delta s_{ba}=s_{ba}-1$. In many circumstances, it is not possible to determine the absolute value of the ensemble average of the conductance, ${\textsl g}$, and in measurements in some geometries, such as optical measurements of transmission in a slab, the precise correspondence of transmission and conductance is not clear. Nonetheless, it is still possible to determine the degree of localization in a sample from measurements of the variances of normalized intensity or total transmission. Because of infinite-range mesoscopic intensity correlation, the fluctuations do not self-average and remain prominent even for the most spatially averaged transmission quantity, the transmittance $T$ \cite{1980a,1980c,1981a,1985b,1985c,1985d,1985e,1998a,2001i,2013k}.

Measurements of the TM for classical waves make it possible to explore long-standing predictions regarding the statistics of conductance in mesoscopic systems in the crossover to localization from the perspective of the transmission eigenvalues \cite{1980a,1997g,1998d,1999f,1999g,1999h,2001f,2002c,2002d}. Unlike measurements of the electronic conductance, for which the measurement of current gives the flux of electrons directly, the transmittance of classical waves is measured on a grid in real space. In such measurements, it is difficult to measure transmission for all incident and outgoing channels. Goetschy and Stone \cite{2013j} have considered the impact of the incomplete measurements of the TM in momentum space upon the statistics of the transmission eigenvalues. They find that the distribution of the density of transmission eigenvalues changes progressively from the bimodal distribution for diffusive waves to a distribution characteristic of Gaussian random matrices \cite{1967a} as the ratio of measured channels $N^\prime$ to the total number of channels $N$ on the input and output sides, $m_1=N_1^\prime/N$ and $m_2=N_2^\prime/N$, decreases. We find the impact of incomplete measurements upon the statistics of eigenvalues is low as long as $N^\prime>{\textsl g}$ \cite{2010h,2013j,2014a}. Thus, the TM for localized waves determined in a single polarization in real space can faithfully represent the statistics of the transmission eigenvalues and the transmittance.

\begin{figure}[htc]
\centering
\includegraphics[width=3in]{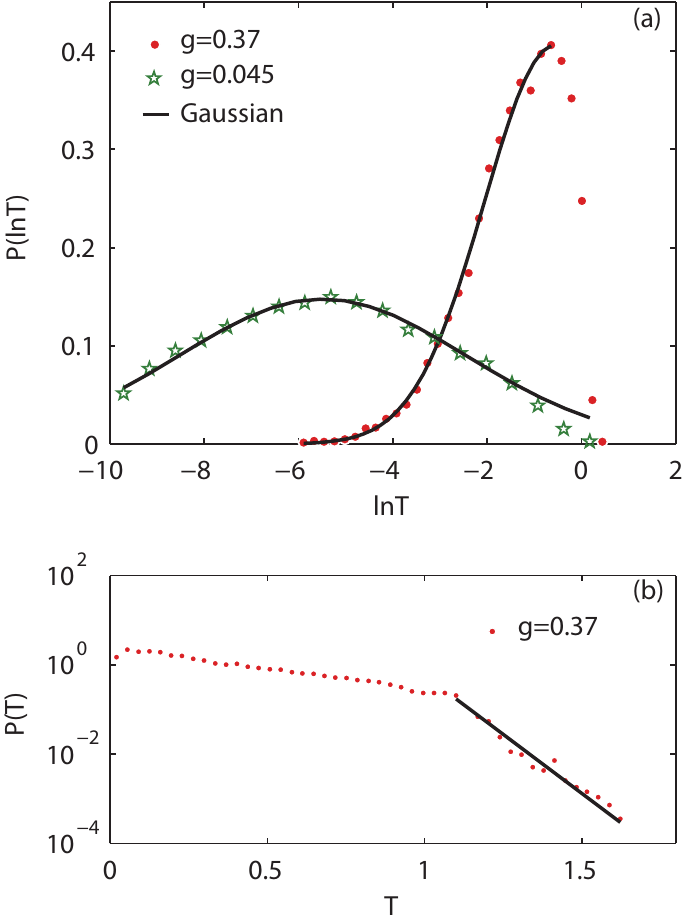}
\caption{Probability distribution of microwave conductance. (a) $P(\ln T)$ for {\textsl g} = 0.37 (red dots) and 0.045 (green asterisk). The solid black line is a Gaussian fit to the data. For {\textsl g} = 0.045, all of the data points are included, whereas for {\textsl g} = 0.37, only data to the left of the peak is used in the fit. (b) $P(T)$ for {\textsl g} = 0.37 in a semi-log plot exhibits an exponential tail. (From Ref. \cite{2014a})} \label{Fig5}
\end{figure}
In Fig. 8, we present the probability distributions of $\ln T$ for microwave radiation transmitted through random ensembles of dielectric samples with ${\textsl g} = 0.37$ and 0.045. Since the measured distributions of the normalized transmittance $s=T/\langle T\rangle$ depends only upon {\textsl g} for localized samples, the value of {\textsl g} for the localized samples can be obtained by comparing the probability distribution $P(s)$ with numerical simulations for a given value of {\textsl g}. $P(\ln T)$ is seen in Fig. 8(a) to be close to a Gaussian distribution for {\textsl g} = 0.045. For {\textsl g} = 0.37, the low-transmission side of $P(\ln T)$ is seen to be well fit by a Gaussian distribution, while the high transmission side falls sharply above the peak at $\ln T=-0.5$. Above $T=1.1$, $P(T)$ is seen in Fig. 8(b) to fall exponentially, in accord with predictions in \cite{2002c}. We show below that in addition to explaining the origin of the distributions of transmittance in the diffusive and localized limits, the Coulomb gas model can explain the anomalous distribution of transmission in the crossover between diffusive and localized transport \cite{1990g,1991c}. Understanding transport in terms of the charge model is of further interest because the charge model provides a parallel treatment of diffusive and localized waves and so forms the basis for a universal description of wave transport.

\begin{figure}[htc]
\centering
\includegraphics[width=3in]{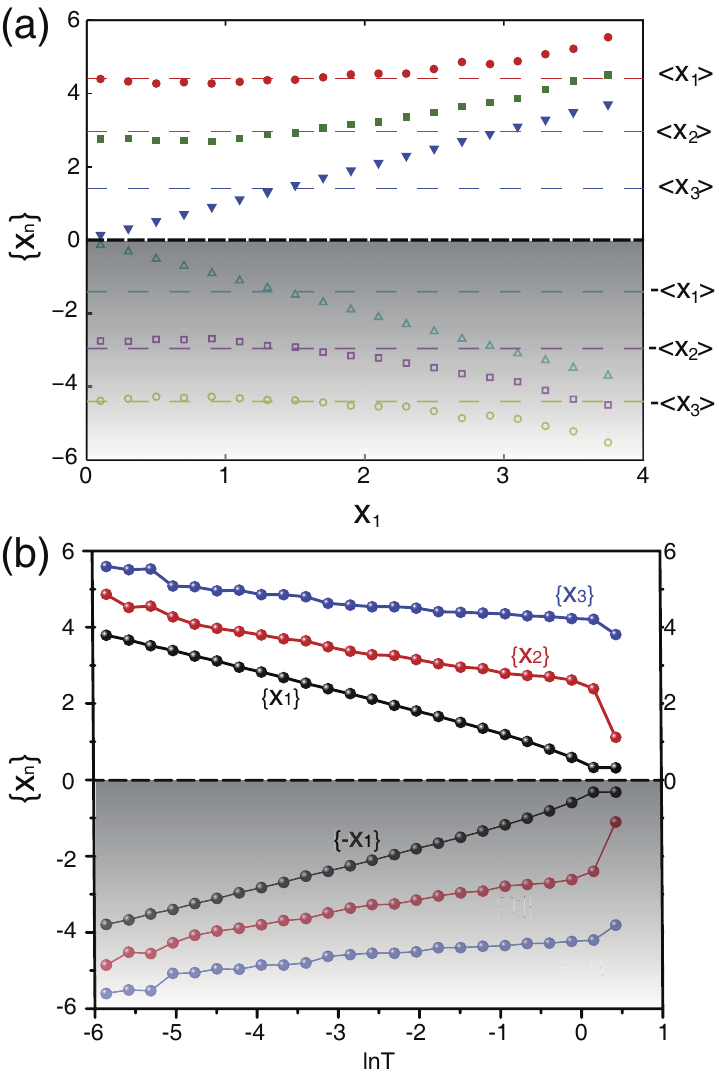}
\caption{Coulomb gas model of transmission eigenvalues and conductance. (a) Average positions of charges and their images as the position of the first charge $x_1$ change in the random ensemble with {\textsl g} = 0.37. The dashed lines show the average positions of the charges for this ensemble. The curly brackets in this case represents the average over a subset of transmission matrices with the specified value of $x_1$. (b) Average positions of charges vs. $\ln T$ in the same ensemble. Here, the curly brackets indicate the averaging is over a subset of transmission matrices with the specified value of $\ln T$. (From Ref. \cite{2014a})} \label{Fig6}
\end{figure}
The variation of the average positions of the charges $x_n$ associated with $\tau_n$ for different positions of the first charge $x_1$ in the random ensemble with \textsl{g}= 0.37 is plotted in Fig. 9. The repulsion between $x_1$, which is associated with the highest transmission eigenvalue $\tau_1$, and its image at -$x_1$ enforces a ceiling for $\tau_1$ of unity. The average spacing between $x_1$ and $x_2$ increases as the value of $x_1$ decreases since the charge at $x_2$ then interacts strongly with the charge at $x_1$ as well as with its nearby image at $-x_1$. At the same time, the spacing between $x_2$ and $x_3$ and their average positions hardly change. This reflects the tendency to heal large fluctuations in charge positions for more remote charges. The source of the sharp cutoff in $P(\ln T)$ can be seen in the context of the Coulomb gas model by examining the spacings of charges for different values of $\ln T$, as shown in Fig. 9(b). A relatively high value of $T$ is only achieved when the first charge is near the origin. This is an unlikely circumstance because this charge is strongly repelled by its image. $P(\ln T)$ would be expected to fall off especially rapidly for values of $T$ above unity since this would requires two charges along with their images to be close together near the origin as seen in Fig. 9(b).

The probability distribution of $\ln T$ for the deeply localized waves is seen to follow a Gaussian distribution. We further explore the relation between the first two moments of the distribution. The ratio of $\sigma^2={\rm var}(\ln T)$ and $-\langle \ln T\rangle$, $\mathscr{R}$, vs. $L/\xi$ is plotted in Fig. 10 and seen to approach unity for $L\gg\xi\sim 24$ cm. 
\begin{figure}[htc]
\centering
\includegraphics[width=3in]{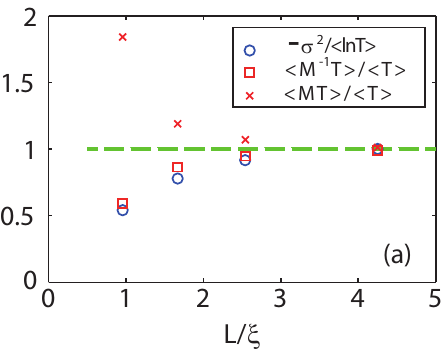}
\caption{Approach to single-parameter scaling in multi-channel random systems. $\mathscr{R}\equiv -{\rm var}(\ln T)/\langle \ln T\rangle$, $\langle M T\rangle/\langle T\rangle$ and $\langle M^{-1} T\rangle/\langle T\rangle$ are plotted vs. $L/\xi$. The dashed line is the prediction of SPS in the limit of deeply localized samples in 1D disordered materials. (From Ref. \cite{2014a})} \label{Fig7}
\end{figure}
This occurs just as the weighted average of $M^{-1}$, $\langle M T\rangle/\langle T\rangle$ and ${\langle M^{-1} T \rangle}/\langle T \rangle $ approach unity. The latter average closely tracks $\mathscr{R}$. Thus, SPS is approached in a multichannel quasi-1D random system when transmission is through a single eigenchannel and transport is essentially one dimensional \cite{1980a,1981d,2001g,2014a}. In this limit, the value of $M$ is close to one in nearly every realization of the random sample except for sample configurations with especially small values of {\it T}, and the relationship between the statistics of transmittance, total transmission and intensity is particularly straightforward. We can therefore obtain the distributions of total transmission and intensity from the distribution of the transmittance by setting $M=1$ and $P(T,M)=P(T)$. Because the distribution of normalized intensity for a given value of $T_a$ is a negative exponential, fluctuations of normalized intensity are closely linked to fluctuations of normalized total transmission, $\langle s_{ba}^n\rangle=n!\langle s_a^n\rangle$. Similarly, in the limit of $M=1$, $\langle s_a^n\rangle=n!\langle s^n\rangle$, where $s=T/\langle T\rangle$. This gives the expression for the variance of $s_{ba}$ when transport is dominated by a single transmission eigenchannel, ${\rm var}(s_{ba})=2{\rm var}(s_a)+1=4{\rm var}(s)+3$ \cite{2014f}. We find the values of the variances of $s_{ba}, s_a$ and {\it s} to be 27.5, 13.2 and 6.4 in the sample of $L$ = 102 cm, which is consistent with these relations. This further confirms that the transport in this random ensemble is via a single eigenchannel. 

\section{Controlling wave propagation in a scattering medium}
Since a random speckle pattern develops on the output plane of a multiply scattering sample illuminated with monochromatic light, it appears that the information regarding incident wave is completely washed out \cite{1975a}. Nonetheless, because the transmitted field is linked to the incident field via the TM of the sample $E_b = \sum_{a}^N t_{ba}E_a$, it is possible to solve the inverse problem and to retrieve the incident waveform by recording the TM and the transmitted field \cite{1981c,1984a,1988c,1997c}. In particular, measuring the TM provides the specific waveform to create a focal spot \cite{2010c,2011h,2012b,2012c,2012d}.

Transmitted waves can be focused at a target point $\beta$ by applying an incident waveform which is the phase conjugation of the field transmission coefficients between the target point $\beta$ and the incident points $a$, $t^*_{\beta a}$ \cite{2008b}. A peak in the intensity emerges at $\beta$, since the transmitted fields from different incident points {\it a} arrive at $\beta$ in phase and interfere constructively while these fields are randomly phased at other points. The incident field is normalized by $\sqrt{\sum_{a}^N |t_{\beta a}|^2}$, so that the incident power is unity. The intensity at focal point $\beta$ is equal to $\sum_{a}^N |t_{\beta a}|^2$, which is the total transmission $T_\beta$ from the point $\beta$ to the opposite surface. This yields a factor of $N$ enhancement over the ensemble average transmission $\langle T_{ba}\rangle$. The background intensity is also increased because of the long range intensity correlation within the TM so that the contrast in focusing is smaller than $N$. The intensity at the focus and in the background can be written as, $I_\beta=\sum_{n}^N \tau_n|u_{n\beta}|^2$ and $I_{b\ne \beta}=\frac{|\sum_n \tau_nu_{nb}u^*_{n\beta}|^2}{T_\beta}$, respectively \cite{2013a}. The contrast for a single measurement of the TM is defined as, $\mu\equiv \langle I_\beta\rangle/\langle I_b\rangle$, in which $\langle \dots \rangle_{b,\beta}$ indicates the average over the background channels $b$ for a given focusing point $\beta$. In the limit of $N\gg 1$, we find,
\begin{equation}
\mu=\frac{1}{1/M-1/N}.
\end{equation}

\begin{figure}[htc]
\centering
\includegraphics[width=2.5in]{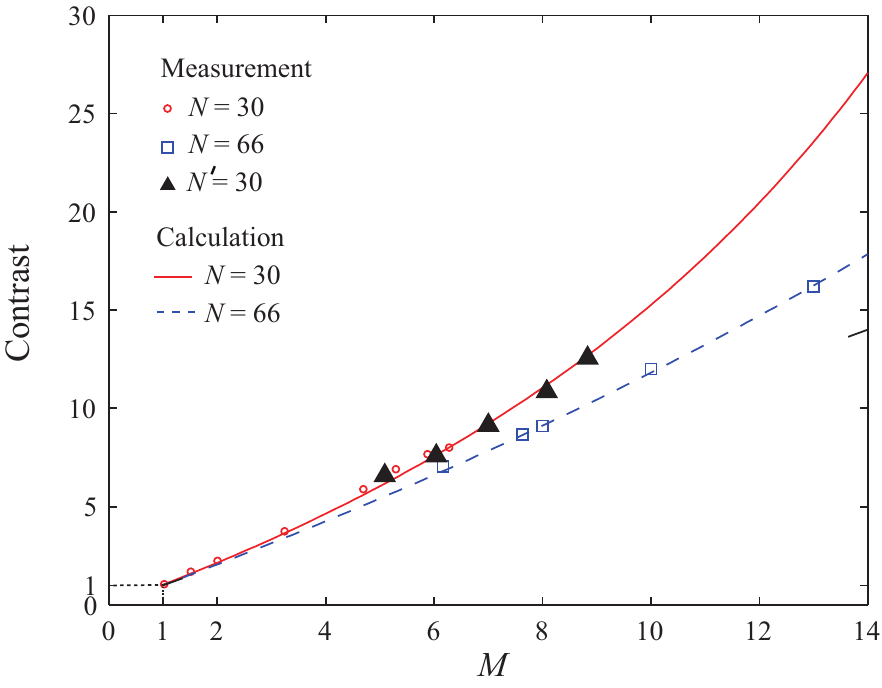}
\caption{Contrast in optimal focusing vs. eigenchannel participation number $M$. The open circles and squares represent measurements from transmission matrices $N$ = 30 and 66 channels, respectively. The filled triangles give results for $N^\prime \times N^\prime$ matrices with $N^\prime$ = 30 for points selected from a larger matrix of size $N = 66$. Phase conjugation is applied within the reduced matrix to achieve maximal focusing. Eq. 4 is represented by the solid red and dashed blue curves for $N =$ 30 and 66, respectively. (From Ref. \cite{2013a})} \label{Fig8}
\end{figure}

The results of measurements of the contrast in optimal focusing for diffusive samples of length $L$ = 23 cm for incident waves in two orthogonal polarization are shown in Fig. 11 to be in excellent agreement with Eq. 4. When the number of measured points $N^\prime$ is smaller than $N$ and therefore the corresponding eigenvalue participation number $M^\prime$ is smaller than $M$ for the complete matrix, the contrast is given by Eq. 4 with the substitutions $M \rightarrow M^\prime$ and $N\rightarrow N^\prime$. This is demonstrated in Fig. 11 with the contrast in samples with {\it N} = 66 but with the contrast computed only for $N^\prime = 30$ points falling on the curve for $N = 30$. These results may be applied to optical measurements of the transmission matrix, where the full matrix is not determined. 

In contrast to a conventional focusing lens, in which the size of the focused beam is diffraction limited and so equal to $\lambda/2$ divided by the numerical aperture of the lens, the profile of the focused beam through a disordered medium depends only upon the property of the random system itself \cite{2002a,2010b,2012d}. The average intensity at a point $b$ at a distance $\Delta r$ from the focal point $\langle I_{foc}(\Delta r)\rangle$ normalized by the peak intensity for a diffusive sample can be expressed in terms of degree of the long range intensity correlation $\kappa$ and the square of the amplitude of the field correlation function, $F(\Delta r)=|E(r)E^*(r+\Delta r)|/\langle I(r)\rangle \langle I(r+\Delta r)\rangle$,
\begin{equation}
\frac{\langle I_{foc}(\Delta r)\rangle_{\beta}}{\langle T_\beta\rangle}=\frac{F(\Delta r)+\kappa}{1+\kappa}.
\end{equation} 
The ratio between the background intensity $F(\Delta r\gg\lambda)$ and the focus intensity $F(\Delta r=0)$ is equal to $\kappa/(1+\kappa)$, since $F(\Delta r)\rightarrow 0$ for $\Delta r\gg \lambda$. In the diffusive limit, $\kappa\sim 1/M\ll 1$, this gives a contrast of $\mu = M$ and the focused profile is reduced to $F(\Delta r)$. Since the intensity correlation function in the diffusive limit is $F(\Delta r)$ according to the field factorization approximation, the profile is then expected to be the same as the intensity correlation, which has been demonstrated in optical measurements \cite{2010b}.
\begin{figure}[htc]
\centering
\includegraphics[width=3in]{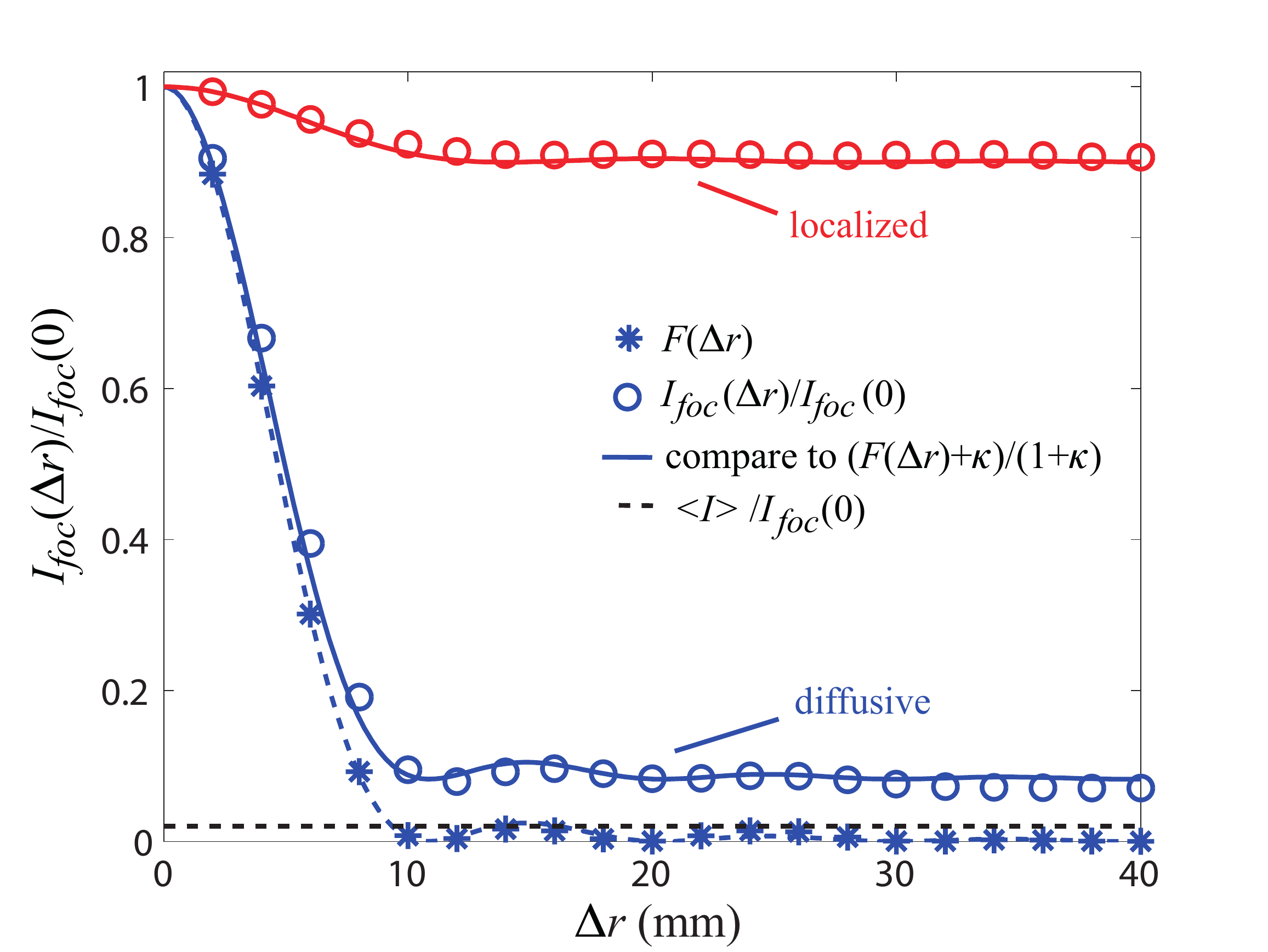}
\caption{The ensemble average of normalized intensity for focused radiation (blue circles) is compared to Eq. 5 (blue solid line) for $L = 61$ cm for diffusive and localized waves. The transmitted field for a single polarization is measured along a line with a spacing of 2 mm and the source antenna is translated along 49 points on a square grid with 9 mm spacing. $F(\Delta r)$ (blue dots) is fit with the theoretical expression obtained from the Fourier transform of the specific intensity (dashed blue line). The black dashed line is proportional to $\langle T_{ba}\rangle/ \langle I_{foc}(\Delta r=0)\rangle = 1/N$. Equation 5 is not valid for localized waves, but good agreement is obtained when $\kappa$ is replaced by $1/(\mu-1)$ in Eq. 5 with $\mu$ determined experimentally. (From Ref. \cite{2012d})} \label{Fig9}
\end{figure}

The normalized profile of the focusing beam via phase conjugation averaged over all possible focusing spot and over all frequency points is presented in Fig. 12. The profile of the focused beam is in excellent agreement with Eq. 5 for diffusive samples. The black dashed line is $1/N$, which is the ratio between the ensemble averaged intensity and the intensity at the focus. The background intensity in optimal focusing is seen to be much greater than the average transmission over a random ensemble and is enhanced by a factor of $N/M$. For localized waves, only a single transmission eigenchannel contributes so that the transmitted wave cannot be focused by applying phase conjugation. 

It is also possible to focus a wave inside a random medium. Since the field inside a material is in general not accessible, many virtual guide stars have been created inside the sample in order to facilitate the focusing at a desired location. These include the use of feedback from florescent particles embedded in the sample \cite{2010g}, photoacoustic \cite{2013h} and acousto-optical effects \cite{2011g,2013e} and the use of many nonlinear phenomena, such second harmonic generation. The contrast of focusing inside a disordered sample has been explored in simulations of a scalar wave propagating through Q1D samples which are locally two dimensional. The wave equation $\triangledown^2E(x,y)+k_0^2\epsilon(x,y)E(x,y)=0$ is discretized on a square grid and solved using the recursive Green’s function method \cite{1991g}. Here, $k_0$ is the wave number in the vacuum and $\epsilon(x,y)$ is the spatial varying dielectric function. $\epsilon(x,y)$ is set to be unity in the two ideal leads and is equal to $1\pm \delta\epsilon$ in the disordered sample. $\delta\epsilon$ is drawn from a rectangle distribution whose width represents the strength of the disorder. Both internal and external reflection are negligible since the sample is index matched to the leads. The product of the wave number in the leads $k_0$ and the grid spacing is unity and the dimension of the samples are measured in units of the grid spacing. The Green's function $G(r,r^\prime)$ is calculated between grid points $r=(0,y)$ on the incident plane and $r^\prime=(x^\prime,y^\prime)$ at depth $x^\prime$ within the interior of the sample with $y$ the transverse coordinate. We considered an ensemble of diffusive samples with length $L$ = 800 and width $L_t$ = 500. The fluctuation of dielectric function at each site is drawn randomly from a rectangular distribution from 0.8 to 1.2. This gives ${\textsl g}=16$.

Optimal focusing at a location at $(x_0, y_0)$ inside the sample can be obtained by phase conjugating the Green's function between arrays of points in the incident plane and the target spot $(x_0, y_0)$. In Fig. 13(a), we present the intensity profile of the focused wave at the center of the transverse dimension $L_t/2$ of the waveguide at two depths $x = L/4$ and $x=L/2$. In both cases, intensity peak emerges at the focal spot with a contrast of $\sim37$ and $\sim17$, respectively. 
\begin{figure}[htc]
\centering
\includegraphics[width=5in]{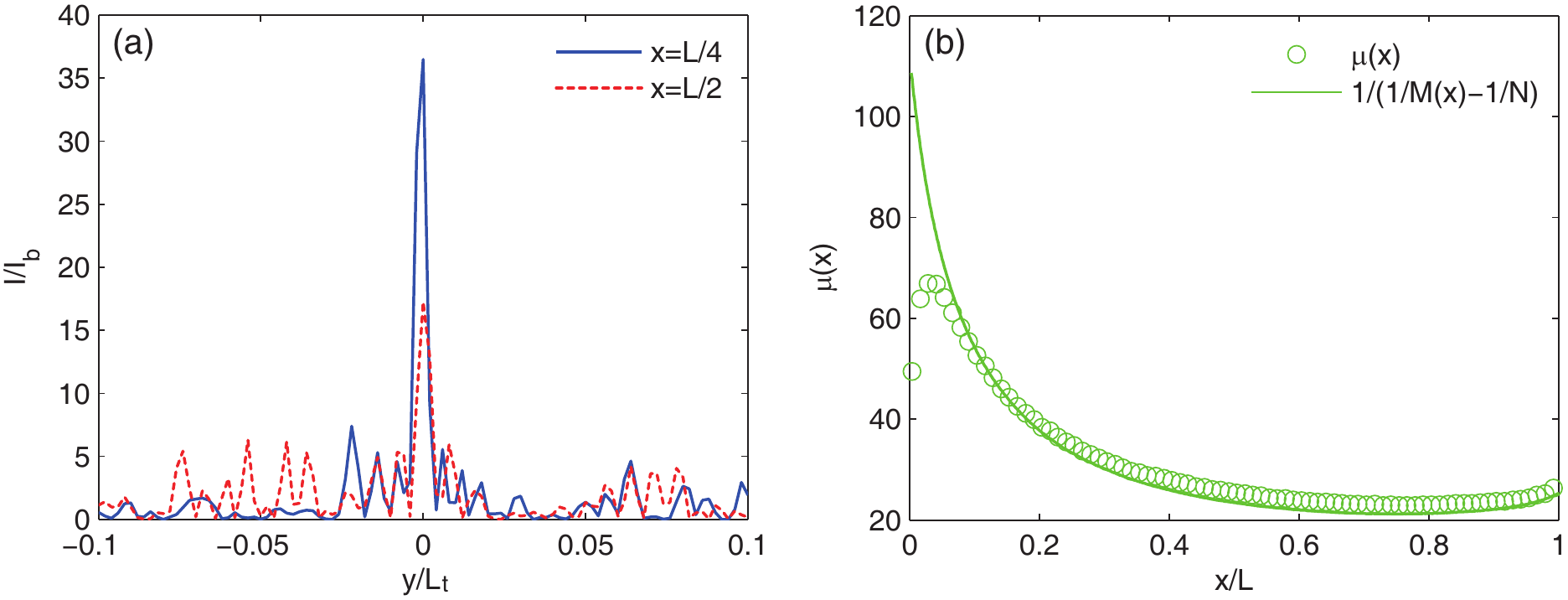}
\caption{Optimal focusing inside a random sample. (a) Intensity distribution in the transverse dimension for optimal focusing at ($L$/4,0) and ($L$/2,0). (b) Spatial variation of the contrast $\mu(x)$ and eigenchannel participation number $M(x)$. (From Ref. \cite{2014b})} \label{Fig10}
\end{figure}

The lower value of contrast for focusing at large depth into the scattering sample is related to the increasing mesoscopic intensity correlation towards the output surface of the sample. The focusing in the sample is given by Eq. 5 with $\kappa$ replaced $\kappa(x)$, the degree of long range intensity correlation at depth $x$. We expect that the contrast of focusing inside a sample is given in terms of the participation number of eigenvalues of the energy density at a depth $x$ into the sample $M(x)$ of the field matrix connecting the field at the incident plane and the focal plane at depth $x$. This hypothesis is confirmed in the plot of spatial variation of the focusing contrast $\mu(x)$ and $1/(1/M(x)-1/N)$ in Fig. 13(b). Good agreement is seen once the depth into the sample is beyond the transport mean free path so that the field is completely randomized. 

When a short pulse is incident on a scattering medium, the pulse is stretched in time as it is scrambled in space. Focusing of a pulse through a random medium was first achieved via the time reversal in acoustics \cite{1992c}. Recently, focusing of a pulse in transmission propagating through an opaque sample has been achieved by adjusting the incident wavefront to enhance transmission at a focal spot at a selected time delay \cite{2011a,2011b,2011k,2013b}. The temporal response of an open random medium to an incident pulse can be naturally described in terms of the quasi-normal-modes of the system \cite{1998e,2006a,2011j}. These modes are characterized by their central frequencies $\omega_m$ and linewidths $\Gamma_m$, and a Lorentzian shape in frequency, $\Gamma_m/2/(\Gamma_m/2+i(\omega-\omega_m))$. Once excited, each of the modes will release its energy at a rate determined by its linewidth $\Gamma_m$, $\exp(-\Gamma_mt)$. Spectra of the transmission coefficient between incident channel $a$ and outgoing channel $b$ can be expressed as a summation of the modes,
\begin{equation}
t_{ba}(\omega)= \sum_{m} t^m_{ba} \frac{\Gamma_m/2}{\Gamma_m/2+i(\omega-\omega_m)}.
\end{equation} 
Here, the matrix $t^m_{ba}$ describes the spatial coupling into the $m^{th}$ mode via incident point {\it a} and coupling out to the output point $b$ from the $m^{th}$ mode. We emphasize that, in contrast to the eigenchannels, the elements $t^m_{ba}$ of two different modes can be strongly correlated \cite{2011j}. The Fourier transform of the spectra of the TM yields a time-dependent TM, which characterizes the spatial-temporal response of the random medium to an incoming pulse, $t(t^\prime)=\sum_m t^m_{ba}\exp(-i\omega_mt^\prime)\exp(-\Gamma_mt^\prime/2)$, where $t^\prime$ indicates the time delay. To have a pulse emerge in transmission at an output point $\beta$ at a certain time delay $t^\prime$, one simply has to apply phase conjugation of the TM at time delay $t^\prime$, $t^*_{\beta a}(t^\prime)$, so that different pulses from incident channels $a$ will be in phase at target point $\beta$ at time delay $t^\prime$. 

\begin{figure}[htc]
\centering
\includegraphics[width=4.5in]{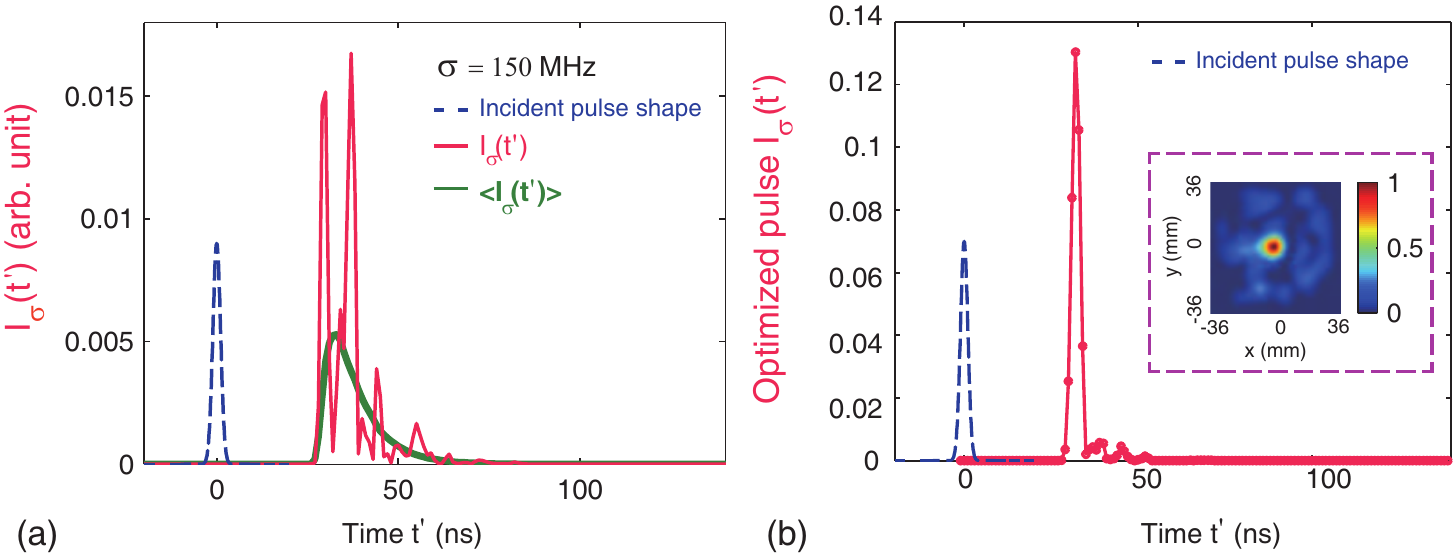}
\caption{Controlling pulse transmission through a random medium with a time dependent TM. Spectra of the TM for microwave radiation propagating through a random waveguide with a length $L = 61$ cm are measured from 14.7 to 15.7 GHz in 3200 steps. The TM is determined with a single polarization between pairs of 45 points on the incident and output surfaces. The time dependent transmission matrix is obtained from spectra of the transmitted field between all points $a$ and $b$, $t_{ba}(\nu)$. These spectra are multiplied by a Gaussian pulse centered in the measured spectrum at $\nu_0=$15.2 GHz with bandwidth $\sigma_\nu=$150 MHz and then Fourier transformed into the time domain. This gives the time response at the detector to an incident Gaussian pulse launched by a source antenna with bandwidth $\sigma_t$=$1/2\pi/\sigma_\nu$. (a) The time variation $I_{ba}(t^\prime)$ of an incident pulse launched at the center of the input surface and detected at the center of the output surface in a single realization of the random sample and the average of the time-of-flight distribution $\langle I(t^\prime)\rangle$. (b) Phase conjugation is applied numerically to the same configuration as in (a) to focus at $t^\prime$= 33 ns at the center of the output surface. The Whittaker-Shannon sampling theorem is used to obtain high-resolution spatial intensity patterns shown in the inset of (b). (From Ref. \cite{2013b})} \label{Fig11}
\end{figure} 
The spatial-temporal control of pulse transmission through a scattering medium via phase conjugation of the time-dependent TM is presented in Fig. 14. The intensity that would be delivered to a point at the center of the output surface of the waveguide $\beta$=(0,0) at a selected time $t^\prime$ if the transmission matrix were phase conjugated at time $t^\prime$ = 33 ns is presented in Figs. 14b. A sharp pulse emerges at the selected time delay with intensity peaked at $\beta$=(0,0). The temporal profile of the focused pulse is the square of the field correlation function in time, $|F_E^\sigma(\Delta t)|^2$, where $F_E^\sigma\equiv \langle E_\sigma(t^\prime)E_\sigma^*(t^\prime+\Delta t)\rangle/(\langle I(t^\prime)\rangle \langle I(t^\prime+\Delta t)\rangle)^{1/2}$. For an incident Gaussian pulse, the square of the field correlation function is equal to the intensity profile of the incident pulse and is independent of delay time. 

The spatial contrast of the focusing of the pulse is given by 
\begin{equation}
\mu=1/(1/M^\prime(t^\prime)-1/N^\prime).
\end{equation}
where $M^\prime$ is the eigenchannel participation number of the TM with a size of $N^\prime$ at time delay $t^\prime$. 
\begin{figure}[htc]
\centering
\includegraphics[width=3in]{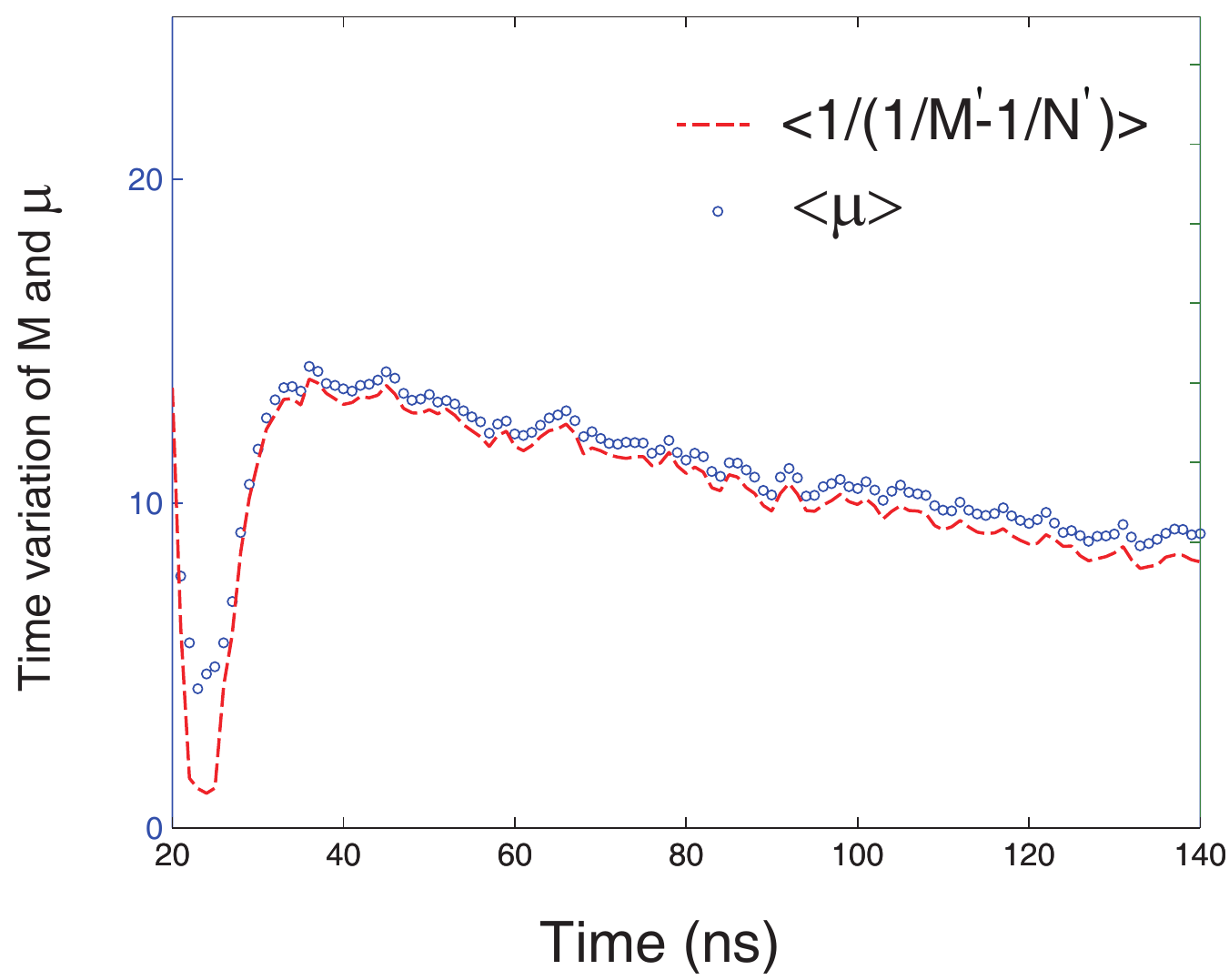}
\caption{Time evolution of the maximal focusing contrast $\langle \mu \rangle$. $\mu$ is well described by Eq. 7 after the time of the ballistic arrival, $t^\prime \sim 21$ ns. At early times, the signal-to-noise ratio is too low to analyze the transmission matrix. (From Ref. \cite{2013b})} \label{Fig13}
\end{figure}
The time evolution of $\langle M^\prime\rangle$ and $\langle \mu\rangle$ are shown in Fig. 15. Near the arrival time of the ballistic wave, the value of $M^\prime$ is close to unity and the contrast is not described by Eq. 7. This is because ballistic wave is associated with the propagating waveguides modes with the highest speed and therefore the transmitted field is not randomized. Once the transmitted wave at the output is multiply scattered, a random speckle pattern develops and the measured contrast is in accord with Eq. 7. After the arrival of ballistic waves, the value of $M^\prime(t^\prime)$ is seen in Fig. 15 to increase rapidly before falling slowly. This reflects the distribution of lifetimes and the degree of correlation in the speckle patterns of quasi-normal modes \cite{2000c,2003a}. Just after the ballistic pulse, transmission is dominated by the shortest-lived modes \cite{1994g,2005b,2006a}. These modes are especially short lived and strongly transmitting because they are extended across the sample as a result of coupling between resonant centers. Sets of extended modes that are close in frequency could be expected to have similar speckle patterns in transmission, so that a number of such modes might then contribute to a single transmission channel \cite{2011j}. As a result, the number of independent eigenchannels of the transmission matrix contributing substantially to transmission would be relatively small at early times and $M^\prime$ would be low. At late times, only the long-lived modes contribute appreciably to the transmission \cite{1983b} and so the value of $M^\prime$ will be close to 1. Thus for intermediate times, modes with wider distribution of lifetimes than at either early or late times contribute to transmission and these modes are less strongly correlated than at early times so that $M^\prime$ and the contrast are peaked.

\section{Dwell time and densities of states of the transmission eigenchannels}
Spectra measurements of the TM allow us to explore the photon dwell time of the transmission eigenchannels. The delay time of transmission in 1D disordered samples is given by $\int I(t)t dt/\int I(t) dt$ \cite{1999e}. In the limit of vanishing bandwidth, it is equal to the derivative of the phase accumulated as the wave propagates through the sample with respect to the angular frequency $\omega$, $\phi^\prime=d\phi/d\omega$ \cite{1999e}. Because the delay time in transmission equals the delay time in reflection, $\phi^\prime$ is also the DOS divided by $\pi$ at angular frequency $\omega$, $\rho(\omega)$. For a multi-channel system, $\phi^\prime_{ba}$ is still the delay time between input channel $a$ and output channel $b$, but it is not loner the DOS. This can be seen by noting that a phase shift of order unity occurs within the field correlation frequency, in which typically a single peak occurs in transmission, but approximately $\delta$ modes contribute to this single peak. Thus the number of modes is much greater than the phase shift. Krein, Birman, Yafaev and Schwinger \cite{1951b,1962b,1992d} have shown that the DOS can be expressed in terms of the scattering matrix {\it S} of the system, $\rho(\omega)=-i/(2\pi)Tr(S^\dagger dS/d\omega)$, in which $-iS^\dagger dS/d\omega$ is known as the Wigner-Smith delay-time matrix \cite{1955a,1960a}. Measurements of the DOS should therefore involve measurements of the complete scattering matrix. For a system with time-reversal symmetry, Brandbyge and Tsukada \cite{1998f} demonstrated that the local DOS in an electronic system can be found from the composite derivative of the phase of the transmitted field with shift in the potential at a point inside the sample. We found that \cite{2015a}, 
\begin{equation}
\rho(\omega)=\frac{1}{\pi}\sum_n^N \frac{d\theta_n}{d\omega}.
\end{equation}
Here, $d\theta_n/d\omega=-i({\bf u_n^*}d{\bf u_n}/d\omega-{\bf v_n^*}d{\bf v_n}/d\omega)$ is eigenchannel density of states (EDOS), which is the contribution to the DOS of the $n^{th}$ eigenchannel and is the dwell time of the $n^{th}$ eigenchannel. $d\theta_n/d\omega$ is thus a complementary set of parameters to $\tau_n$. Thus spectra of the TM yield the dynamics of transmission as well as static transmission for each eigenchannel. 

\begin{figure}[htc]
\centering
\includegraphics[width=5in]{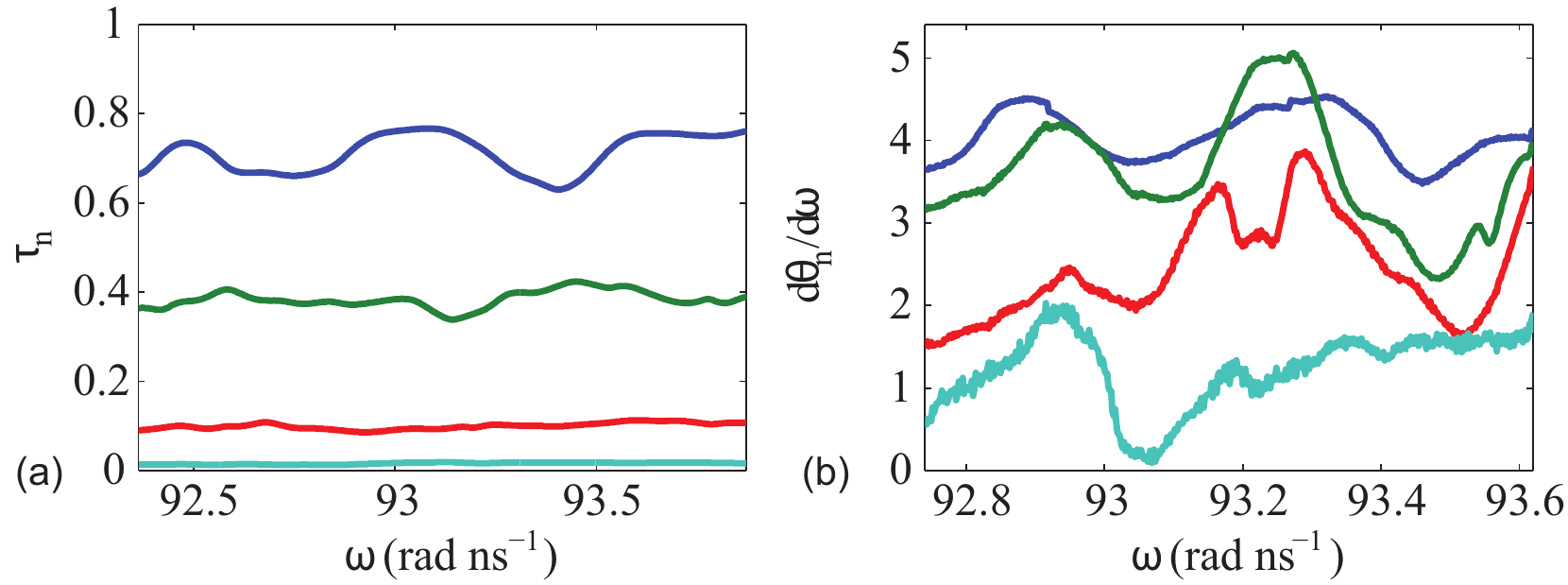}
\caption{Spectra of transmission eigenvalues $\tau_n$ and the eigenchannel dwell times $d\theta_n/d\omega$ for {\it n} = 1, 5, 15, 25 for a diffusive sample with {\textsl g} = 6.9. The dwell time for the eigenchannels is the contribution of the eigenchannel to the DOS. (From Ref. \cite{2015a})} \label{Fig14}
\end{figure}
Spectra of $\tau_n$ and $d\theta_n/d\omega$ for a diffusive sample with ${\textsl g}=6.9$ are presented in Fig. 16. The average value of $d\theta_n/d\omega$ is seen to decrease as the eigenchannel channel index {\it n} increases, indicating that the average dwell time becomes shorter as the value of transmission decreases. The sum of $d\theta_n/d\omega$ represents the DOS of the sample, when the impact of incompleteness of the TM is low. We therefore expect that Eq. 8 gives the DOS in samples for which $N^\prime \gg M$, as is readily achieved in samples in which the wave is localized. The DOS of the sample can also be found by identifying and counting the contribution of the each quasi-normal mode inside the sample. For an open system, the DOS can be expressed as,
\begin{equation}
\rho(\omega)=\sum_m \rho_m(\omega)=\frac{1}{\pi}\frac{\Gamma_m/2}{(\Gamma_m/2)^2+(\omega-\omega_m)^2}.
\end{equation}
The central frequencies and linewidths of the modes have been found for localized samples where the modal overlap is small, by decomposing the transmitted field speckle pattern into a sum of the same set of modes \cite{2011j}. This allows us to directly measure the DOS of the disordered sample. Good agreement based upon these two approaches for the localized sample with {\textsl g} = 0.37 is seen and demonstrated in Fig. 17.
\begin{figure}[htc]
\centering
\includegraphics[width=5in]{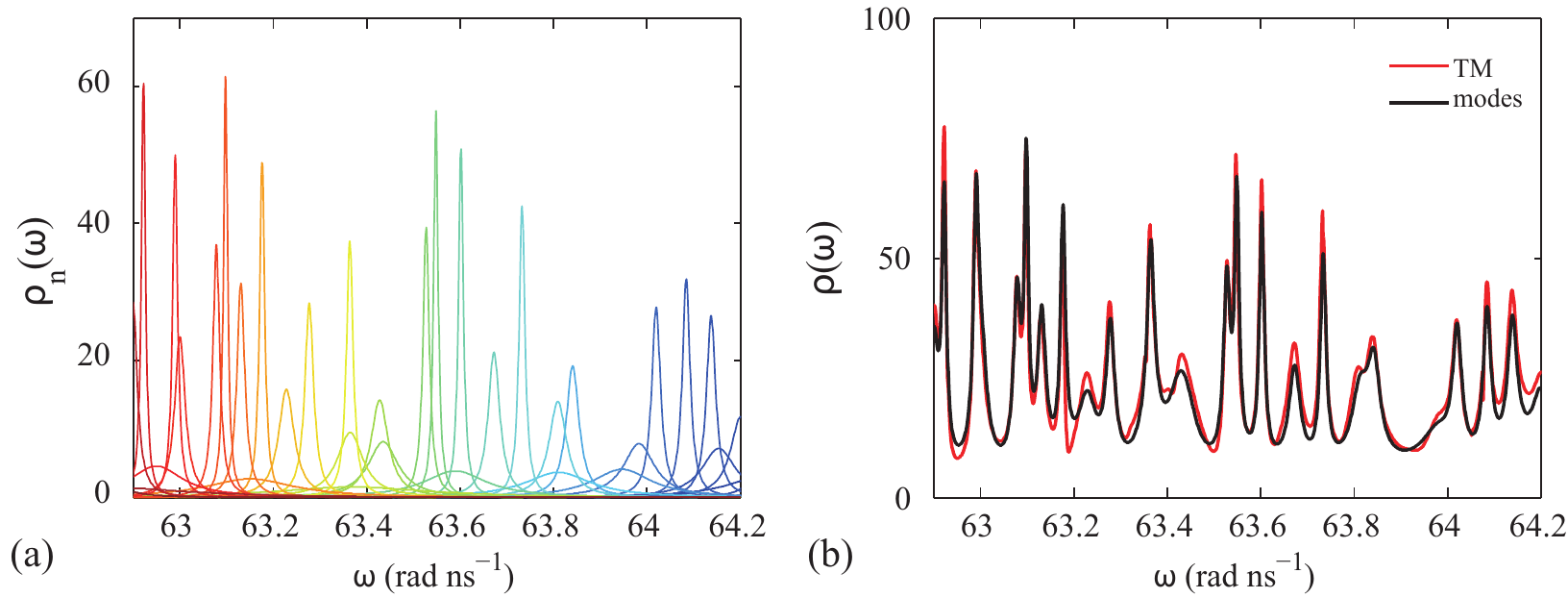}
\caption{Comparison of two approaches for finding the DOS of a disordered material. (a) Contributions of the individual modes in Eq. 9 to the DOS. The integral of each mode over the angular frequency is unity. (b) DOS determined from the TM (red curve) by summing spectra of $d\theta_n/d\omega$ and modes in (a). (From Ref. \cite{2015a})} \label{Fig14}
\end{figure}

Since the photon dwell time and the EDOS of an eigenchannel are proportional to the energy stored inside the sample \cite{1995d}, this suggests that the energy stored inside is smaller for low transmitting eigenchannels. This is consistent with the results found simulations by Choi {\it et. al.,} \cite{2011e} and measurements by G\'{e}rardin {\it et. al.} \cite{2014d} in single samples. An expression of the profile of an eigenchannel within a random medium would be of importance, since it would afford a universal description for wave propagation that encompasses the wave inside the sample as well as the transmitted wave. This would also provide preliminary sense of the intensity inside individual eigenchannels channels for a host of application involving energy deposition inside random media. This may also be exploited to lower the lasing threshold of a diffusive random laser \cite{AA,BB}, in which the threshold is particularly high because the pump energy cannot penetrate deep into the sample \cite{1994e}. The profiles of eigenchannels over a wide range of transmission obtained in 2D simulations using the recursive Green's function method are shown in Fig. 18. The profiles are averaged over a collection of samples with the same transmission $\tau$. The profile of the complete transmission eigenchannel with $\tau = 1$ is found to be closely related to the probability of a wave returning to a cross-section at a depth {\it x} in an open disordered medium \cite{2008c,2008d,2010e,2015b}. We have found that the profile of an eigenchannel with $\tau<1$ can be expressed as a product of the profile of the fully transmitting eigenchannel and a function governed only by the auxiliary localization length $\xi^\prime$, which was previously used to parameterize the corresponding transmission eigenvalues via $\tau=1/\cosh^2(L/\xi^\prime)$. These results not only give a physical meaning to the set of $\xi^\prime$ but also enable control of the energy deposition inside the opaque media. 
\begin{figure}[htc]
\centering
\includegraphics[width=5in]{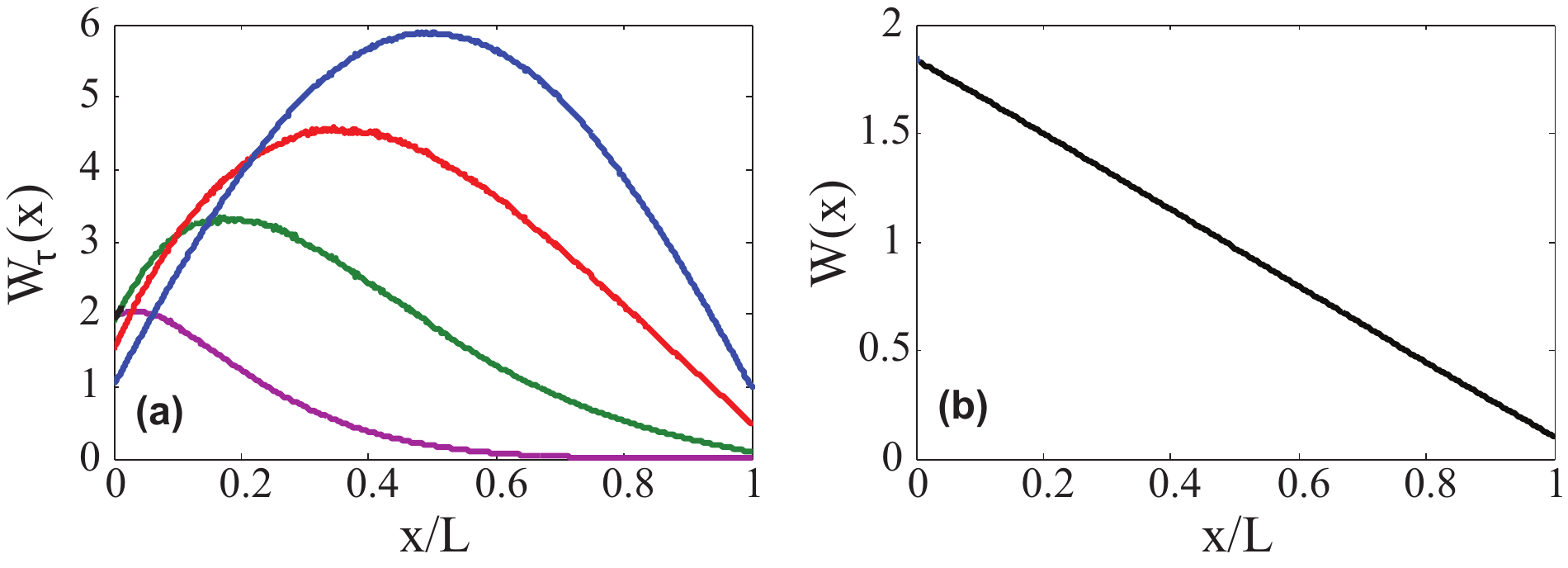}
\caption{Spatial profile of the energy distribution of transmission eigenchannel inside the sample. (a) Ensemble averages of the eigenchannel energy density profiles $W_\tau(x)$ for eigenvalues $\tau=1$, 0.5, 0.1, and 0.001 for a diffusive sample with $L/\xi=0.05$. $W_\tau(x)$ is the energy integrated over the transverse dimension and is normalized to equal $\tau$ at the output surface. (b) Ensemble averages of energy density profiles for all transmission eigenchannels with the eigenchannel indices {\it n} from 1 to {\it N}. The linear falloff of the average of the energy density over all eigenchannels is in accord with the diffusion theory. (From Ref. \cite{2015b})} \label{Fig16}
\end{figure}

\section{Conclusions}
In this review, we have discussed the power of the transmission matrix to describe the statistics of transmission and control of waves propagating through disordered systems. We have shown that the joint probability distribution $P(T,M)$ determines the statistics of transmission over a random ensemble up to second order, while the statistics within a single sample of given $T$ depends only on $M$. The statistics of transmission in multi-channel disordered systems approach predictions for 1D, when the transmission is dominated by a single transmission eigenchannel and so $M\rightarrow1$. In this limit, the SPS hypothesis for scaling in random 1D system holds for multi-channel disordered samples. The contrast of optimal focusing through and within a scattering sample is equal to $M$ of the measured TM, provided that the number of the measured channels of the TM is much greater than {\textsl g}. The spatial profile of the focused wave is given in terms of the degree of long-range intensity correlation and the spatial field correlation function while the temporal profile of a focused pulse is equal to the square of the field-field correlation function in time. In addition to the transmission eigenvalues, which describe static transmission, there exists a complementary set of parameters, which gives the dwell time and the integrated intensity inside the sample for transmission eigenchannels, and the contribution of each eigenchannel to the DOS. The TM approach allows us to determine the DOS of a disordered material without finding each of the QNMs in the system, which can be difficult for diffusive samples because of the modal overlap. The ability to deposit energy deep within the sample via high-transmission eigenchannels may find applications in deep tissue imaging, depth profiling, photodynamic therapy and random lasing. 

\section{Acknowledgments}
We thank Chushun Tian, Xiaojun Cheng, and Jerome Klosner for stimulating discussions and Howard Rose for construction of the microwave sample holder. The research was supported by the National Science Foundation (DMR-1207446).

\end{document}